\documentstyle[epsfig]{mn2e}
\begin{document}

\title
[Dwarf galaxies in the Dynamically Evolved NGC 1407 Group]
{Dwarf galaxies in the Dynamically Evolved  NGC 1407 Group}

\author[Neil Trentham et al.]
{
Neil Trentham$^{1}$, R.~Brent Tully$^{2}$ and Andisheh Mahdavi$^{2,3}$\\
$^1$ Institute of Astronomy, Madingley Road, Cambridge, CB3 0HA.\\
$^2$ Institute for Astronomy, University of Hawaii,
2680 Woodlawn Drive, Honolulu HI 96822, U.~S.~A.\\
$^3$ Physics and Astronomy, University of Victoria,
Victoria, BC V8W 3P6, Canada \\
}
\maketitle

\begin{abstract} 
{ The NGC~1407 Group stands out among nearby structures by its
properties that suggest it is massive and evolved.  It shares
properties with entities that have been called fossil groups: the
$1.4^m$ differential between the dominant elliptical galaxy and the
second brightest galaxy comes close to satisfying the definition that
has been used to define the fossil class. There are few intermediate
luminosity galaxies, but a large number of dwarfs in the group. We
estimate there are 250 group members to the depth of our survey.  The
slope of the faint end of the luminosity function (reaching $M_R =
-12$) is $\alpha = -1.35$.  Velocities for 35 galaxies demonstrate
that this group with one dominant galaxy has a mass of $7 \times
10^{13} M_{\odot}$ and $M/L_R = 340 M_{\odot}/L_{\odot}$.  Two
galaxies in close proximity to NGC~1407 have very large blueshifts.
The most notable is the second brightest galaxy, NGC~1400, with a
velocity of $-1072$~km~s$^{-1}$ with respect to the group mean.  We
report the detection of X-ray emission from this galaxy and from the
group.}
\end{abstract} 


\section{Introduction} 

This paper presents a continuation of a study of the dwarf galaxy 
populations in a variety of environments within the Local Supercluster
(Trentham, Tully, \& Verheijen 2001; Trentham \& Tully 2002 [TT02]; 
Mahdavi, Trentham, \& Tully 2005 [MTT05]).  Our goals are to give better
precision to the properties of the luminosity function of galaxies at 
faint levels and to determine the nature of any variations
with environment.  

Our earlier observations affirmed the claims of a
deficiency of low-luminosity dwarf galaxies compared with the large
number of low-mass dark-matter halos predicted by cold dark matter
(CDM) theory (Klypin et al. 1999, Moore et al. 1999).  The discrepancy
is large -- more than one order of magnitude -- and potentially a
problem for CDM theory.  These studies also point towards a
situation where there are more low-luminosity dwarf galaxies per
high-luminosity giant galaxy in dense clusters than in the field.
This discrepancy is much weaker: studies of the Virgo cluster
(e.g.~Trentham and Hodgkin 2002) showed that there are about twice as
many dwarf galaxies per giant there than in the field.  There is 
an upturn in the cluster luminosity function at $M_B = -16$ that is
not seen in the field luminosity function.  Interestingly this upturn
has also been seen in a large study of rich clusters using SDSS data
(Popessa et al.~2005).  The Virgo data is deep enough and the
cluster is near enough that we can see the luminosity function begin
to flatten again at about $M_B = -13$.

There are a number of physical processes during galaxy formation which
could generate the observed phenomena.  For example, the suppression
of the formation of low-mass galaxies may be a strong function of the
state of the intergalactic medium.  At later times, after
reionization, the intergalactic medium is too hot to permit the
collapse of gas into any small dark-matter halos that may form.
If gas must be cold to accumulate, there will be more dwarfs per giant 
in any part of the
Universe where galaxies form early, for example, galaxy clusters.
Another mechanism might be the pressure-confinement of gas within
galaxies by an intracluster medium.  This would reduce the feedback
effects of supernova-driven winds in low-mass galaxies in galaxy
clusters, where such a medium is present.  Yet another mechanism would
be the inability of giant galaxies to consume dwarfs in galaxy
clusters, where the velocity dispersion is high.

Simulations of galaxy formation that investigate the roles of these
different physical processes are limited by a lack of observational
data, and it is this deficiency that our study attempts to address.  
Our plan
is to measure the galaxy luminosity function to faint limits in
environments that have had different formation histories.  The
availability of large format digital detectors provides an opportunity
to gather much better statistics on the occurrence of dwarf galaxies.
The present study is based on observations with the
Canada-France-Hawaii Telescope 12K detector, the predecessor of the
current MegaCam CCD mosaic (Aune et al.~2003).  The ultimate intention 
is to survey a
representative sample of nearby groups and clusters that span wide
ranges of density and mass.  MTT05 presented results on the first
region to be fully surveyed in our program, the intermediate mass but
dense NGC~5846 Group.

Small, dense groups were observed earliest in the program
for the practical reason that they subtend modest areas on the sky. More
extended groups have subsequently been observed with the larger format
MegaCam at CFHT and the results from these studies will follow.  The
current study concerns an unusual group in the context of the Local
Supercluster region.  The group of galaxies centered on NGC~1407 is
sparse in luminous systems, but its velocity dispersion is high, and
the group appears to be dynamically evolved.  It shares the properties
of `fossil' groups (Ponman et al. 1994).  It fails slightly to meet
the formal qualification for such a group that there be a
differential of at least 2 magnitudes between the brightest and second
brightest objects since the interval between NGC~1407 and 
NGC~1400 is only 1.4 magnitudes.
In any event, among structures with $V <2,000$~km~s$^{-1}$ the 
NGC~1407 Group is extreme.

\section{The Unusual NGC~1407 Group}

This small group of E and S0 galaxies dominated by NGC~1407 has
attracted attention because of a suspected high mass-to-light ratio
$M/L$ (Gould 1993; Quintana, Fouqu\'e, and Way 1994).  The group
velocity dispersion based on 35 measurements is $387 \pm
65$~km~s$^{-1}$.  The mass of the group found from application of the
virial theorem is $7.3 \times 10^{13} M_{\odot}$.  The mass-to-light
ratio in the $R$ band is then $M/L_R = 340 M_{\odot}/L_{\odot}$.
Astonishingly, the second brightest galaxy in the group, NGC~1400, is
{\it blueshifted} by $1072$~km~s$^{-1}$ with respect to the group mean of
1630~km~s$^{-1}$, a line-of-sight velocity excursion of $2.8 \sigma$.
If the 3D velocity dispersion is proportional to the projected
excursion and mass is distributed like the light, then 75 per cent of
the orbital kinetic energy of the group would be attributed to this
one galaxy.  This circumstance is unlikely.  The most plausible
resolution is that most of the mass is in an extended group halo and
that NGC~1400 is akin to a test particle of negligible mass, presently
near the potential minimum and directed close to our line of sight.
Surface brightness fluctuation measurements (Tonry et al. 2001) assure
that NGC~1400 is indeed at the same distance as NGC~1407.  Note that
this reasonably direct inference that most of the group mass is
decoupled from the light strongly negates the hypothesis of modified
Newtonian gravity (Milgrom 1983).

The existence of groups with such high mass-to-light 
ratios is reasonable in the context of $\Lambda$CDM cosmology (see Gao et al.~2004 for
a description of some of the important physical processes).
That the group around NGC 1407 is an example of such a group comes mainly from dynamical evidence,
of which the relative velocity of NGC 1407 and NGC 1400 is the most striking.
Given that surface brightness fluctuations place these two early-type galaxies at the same distance, the probability
of their {\it not} being bound is extremely small.
Nevertheless, it is important to recognize the alternative (although we believe less likely) possibility that NGC 1400 
has a high velocity relative to NGC 1407 because it is passing through the group, maybe for the first time.

The NGC~1407 Group resembles but is more extreme than the NGC~5846
Group discussed by MTT05.  The NGC~5846 Group is a dumbbell system of
$8 \times 10^{13} M_{\odot}$ arranged about two dominant ellipticals.
The NGC~1407 Group has essentially as much mass but with a tighter
concentration around the central large elliptical 
and an even greater predominance of early-type galaxies.

These characteristics lead us to identify the
structure as akin to a "fossil group", an entity at an advanced
dynamical stage.  After a
search for small groups dominated by an elliptical at least 2
magnitudes brighter than any companion, Khosroshahi et al. (2004)
identified the group around NGC~6482 at $V_{GSR} = 4089$~km~s$^{-1}$
as the nearest entity to satisfy the `fossil' criterion.  The NGC~1407
Group at $V_{GSR} = 1546$~km~s$^{-1}$ does not quite satisfy the
magnitude differential criterion but it is the most extreme structure
of a similar nature within 2000~km~s$^{-1}$.
Another difference between this group and fossil groups concerns 
large-scale environment; the NGC 1407 
Group exists in a richer supercluster environment than most fossil 
groups.
Given these differences it would not be correct to refer to this group
as a fossil group.
Nevertheless it is intriguing that such a nearby group shares so many 
properties with these rare objects.

\begin{table} \caption{Properties of the NGC 1407 Group}
{\vskip 0.75mm}
{$$\vbox{
\halign {\hfil #\hfil && \quad \hfil #\hfil \cr
\noalign{\hrule \medskip}
\cr
Designation$^{*}$ &51 $-$8&\cr
Distance (Mpc)     & 25 &\cr
No.~E/S0/Sab~$M_R<-19$ &13  &\cr
No.~Sb-Irr~$M_R<-19$   & 1 &\cr
Velocity dispersion (km s$^{-1}$) & 387 &\cr
Inertial radius (Mpc) & 0.39 &\cr
Crossing time & 0.07 $H_0^{-1}$ &\cr
Log$_{10}$ (R luminosity/L$_{\odot})$
                     & 11.33 &\cr
Log$_{10}$ (Mass/M$_{\odot})$
                     & 13.87 &\cr
R band mass-to-light ratio / M$_{\odot}$/L$_{\odot}$
                     & 340 &\cr
Density/Mpc$^2$ at 200 kpc $M_R<-17$
                     & 34 &\cr
Dwarfs / giants      & $6.5\pm1.3$ &\cr
\noalign{\smallskip \hrule}
\noalign{\smallskip}\cr}}$$}
\begin{list}{}{}
\item[$^{*}$]This is the designation given in the {\it Nearby Galaxies
Catalog} (Tully 1988).
\end{list}
\end{table}

\subsection{The group environment}

Before studying the group properties, we need to look in some detail
at the region of sky around the group because of concerns regarding
projection contamination.  The NGC~1407 Group, 51-8 in the Nearby
Galaxies Catalog (Tully 1988), is 18 degrees ($\sim 8$~Mpc) away in
projection from the Fornax Cluster (51-1).  It is part of the Eridanus
Cloud which contains other knots of early type systems in the
vicinity.  The extended region called 51-4 containing the dominant 
galaxies NGC 1332, 1371, 1395, and 1398 is 2--3~Mpc removed in projection
and 300~km~s$^{-1}$ blueshifted in the mean.  The Eridanus region
excluding the NGC~1407 Group must contain at least $3 \times 10^{13}
M_{\odot}$ (the combination of groups 51-4,5,7 in Table 1 of Tully
2005).  Though the NGC~1407 Group appears distinct today it must be
strongly suspected that the entire core of the Eridanus Cloud is bound
with over $10^{14} M_{\odot}$ and will ultimately merge.

The first two figures provide information about the group environment.
Figure~1 is a histogram of all known velocities within
10,000~km~s$^{-1}$ in the projected area shown in Figure~2.
The histogram reveals that two structures, the Eridanus Cloud and the
Southern Wall, dominate the field in the region of interest.
There is no known contamination foreground of the Eridanus Cloud and
there is a void behind Eridanus that extends to the Southern Wall.
The Southern Wall is 2.5 times more distant so its members are about 2
mag fainter than the equivalent galaxies in Eridanus.

\begin{figure}
\begin{center}
\vskip-4mm
\psfig{file=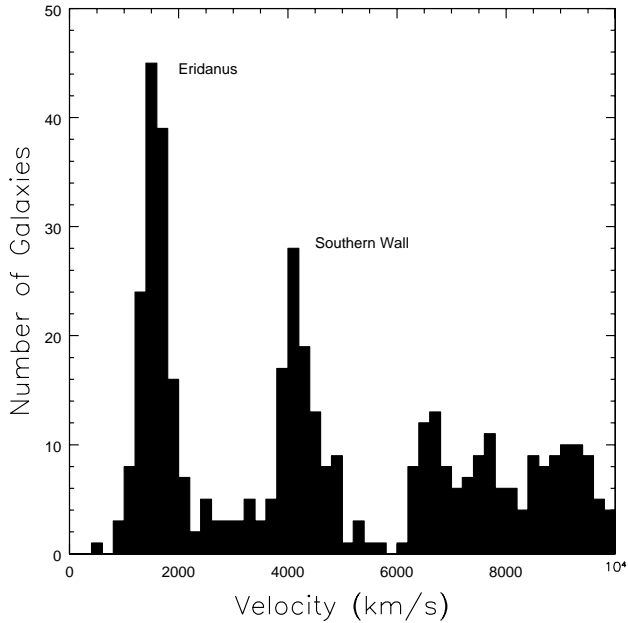, width=8.65cm}
\end{center}
\vskip-3mm
\caption{     
Histogram of velocities of galaxies projected onto the Eridanus region
shown in Fig.~2.
}
\end{figure}


\begin{figure*}
\begin{center}
\resizebox{7in}{!}{\includegraphics{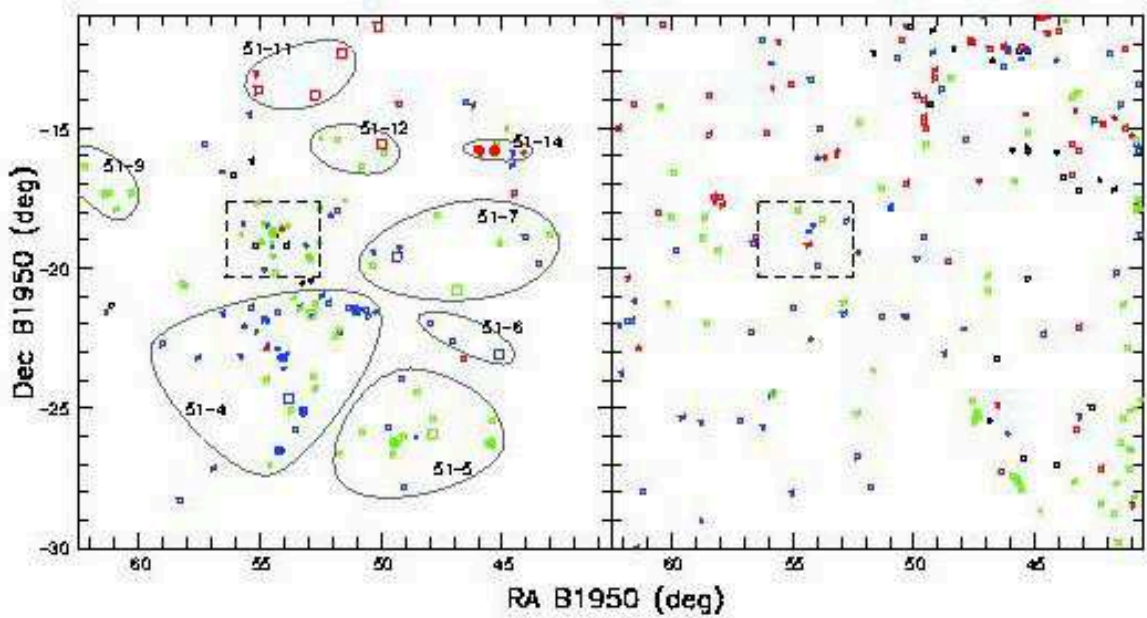}}
\caption{
The left panel shows the distribution of galaxies within the velocity range of the
Eridanus Cloud ($v<$ 3000~km~s$^{-1}$.  Filled round symbols 
refer to early-type ($T \leq 1$) galaxies.  Open square symbols 
refer to late-type ($T > 1$) galaxies.  More luminous galaxies have
larger symbols.  Our CFHT survey region around NGC~1407 is indicated by the 
small rectangle.  The irregular enclosed regions indicate the locations
of the main groups identified in the Nearby Galaxies Catalog.  The
51-4, 51-6, and 51-7 groups have somewhat lower mean velocities than
the 51-8 (NGC~1407) group, the 51-5 and 51-9 groups have comparable 
mean velocities, and the 51-11, 51-12, and 51-14 groups have higher
mean velocities.
The right panel shows the distribution of galaxies background of the 
Eridanus Cloud out to 10,000~km~s$^{-1}$.
\label{fig:erimap}}
\end{center}
\end{figure*}

The two panels of Fig.~2 show the distribution of galaxies with known
velocities in the Eridanus Cloud and the near-background.  The 51--4 group
containing NGC~1395 is seen to be nearby but distinct.  The NGC~1407
Group is contiguous with the other components of the Eridanus Cloud
but there is no suggestion of confusion.  Overwhelmingly, candidates
within the survey region are early types but most galaxies outside a
few discrete knots in Eridanus are late morphological types, very few
of which are projected into the survey region.

We do, however, note an unfortunate instance of projection.  Three S0
galaxies (NGC~1394 the brightest) in a knot within the Southern Wall
lie near the middle of the CFHT survey region.  These galaxies are
clearly distinguished as background by their velocities, and are not
part of the Eridanus Cloud described in the previous paragraph.  In
the current study, we must give consideration to the possibility that
low-surface-brightness companions to these systems could enter our
list of NGC~1407 Group candidates.  More distant galaxies are expected
to be excluded from our candidate list by the procedures described in
section 4.

\subsection{X-ray properties}

The X-ray-emitting medium in clusters of galaxies is an important
tracer of the overall matter distribution. Unfortunately, in groups of
galaxies the surface brightness is so low relative to the X-ray
background that the emission can rarely be detected beyond a few
tenths of a virial radius. The NGC 1407 galaxy was observed by the
Einstein (Dressler \& Wilson 1985), ROSAT (Davis \& White 1996), ASCA
(Buote \& Fabian 1998), and Chandra (Zhang \& Xu 2004) X-ray
telescopes. The Chandra observation, which has the best spatial
resolution, shows a diffuse X-ray-emitting medium 
with temperature 0.7 keV and metal abundance about
0.4 solar, a weak central AGN
and 40 X-ray point sources, most of which are X-ray
binaries. The X-ray luminosity of the diffuse
component is $4 \times 10^{40}$ erg s$^{-1}$ in the 0.3--10 keV
band. The NGC 1407 Group is then underluminous by
an order of magnitude given the expected value from the X-ray
luminosity-temperature and luminosity-velocity dispersion relations
(Mahdavi \& Geller 2001, Osmond \& Ponman 2004). 

Because the system is so nearby, the ASCA and Chandra observations of
NGC 1407 include little of the group-scale emission outside the
central galaxy.  The wider
field of the ROSAT PSPC detector was able to capture some of this
group-scale emission with a resolution of $\approx 15\arcsec$
(see Figure 3). 
Osmond \& Ponman (2004) use a $\beta$-model 
to determine a total X-ray luminosity of $\approx 10^{42}$ erg s$^{-1}$ within
570 kpc, which puts the NGC 1407 group directly on the X-ray
luminosity-temperature and luminosity-velocity dispersion relations
(rather than making it underluminous, as do the Chandra data). One
caveat is that an extrapolated
$\beta$-model can overestimate the gas density at the largest radii
(Markevitch et al. 1998).
It seems that the NGC 1407
Group is not an X-ray \emph{overluminous} system of galaxies. This is
in contrast to the average fossil group, which is overluminous in the
X-ray relative to a non-fossil counterpart of the same mass
(Khosroshahi et al. 2004; D'Onghia et al. 2005). 

The ROSAT data also show significant X-ray emission associated with
NGC 1400 and with the probable member dwarf galaxy LEDA
074845. 
Fitting a MEKAL thermal plasma model to these peaks, we find that the emission in
both the peaks comes from gas with the same
temperature and metallicity 
as the gas around NGC 1407. 
The resulting 0.1-10 keV X-ray fluxes are $1.5
\pm 0.3 \times 10^{-13}$ ergs s$^{-1}$ cm$^{-2}$ for NGC 1400 and $1.1 \pm 0.4
\times 10^{-13}$ erg s$^{-1}$ cm$^{-2}$ for LEDA 074845. We have
submitted an XMM-Newton proposal to further investigate the nature of
the puzzling emission near LEDA 074845.
To our knowledge this is the second reported detection of X-ray
emission from NGC 1400---other reports, e.g. Canizares et al.~(1987)
and Beuing et al.~(1999) have been upper limits
but O'Sullivan, Forbes and Ponman (2001) report a detection of this galaxy.

Outside the region shown in Figure 3, the ROSAT All Sky Survey shows a
lack of X-ray emission from galaxies associated with the NGC 1407
Group (or any other group). Within 0.9 Mpc ($2 \degr$) of NGC 1407,
there are no galaxies, groups, or clusters with $c z < 20,000$ km
s$^{-1}$ that have detectable X-ray emission.

\begin{figure}
\begin{center}
\vskip-4mm
\psfig{file=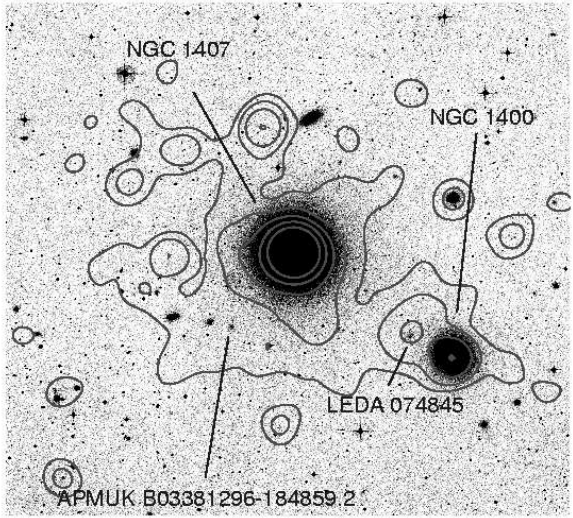, width=8.65cm}
\end{center}
\vskip-3mm
\caption{ ROSAT X-ray flux contours superposed on an image of the
central $35 \arcmin \times 30\arcmin$ region of the NGC~1407 Group.
The outermost contours show emission at 2$\sigma$ significance; each
successive contour represents an enhancement in the significance by a
factor of 1.5.}
\end{figure}

\section{Observations and data reduction}

\subsection{Imaging observations}

Observations were made with the Canada-France-Hawaii Telescope (CFHT) 12K CCD
camera in queue mode over 6 photometric nights between 8 November and 
6 December, 2002.
Seeing was 0.6--0.8 arcseconds FWHM.
The CFHT 12K camera (Cuillandre et al.~2001) is a mosaic of 12 CCD
detectors providing a field of $42^{\prime} \times 28^{\prime}$,
oriented in this program with the long axis E-W.  The observations
were made with half-field overlaps so that gaps between
CCD chips in any particular exposure were almost entirely filled in and
most of the area was observed twice,
therefore allowing any non-photometric data to be calibrated.  
In total, 69 x 11 minute
exposures were taken, all in the Cousins R band, covering 10.05 square
degrees. 
A rectangular area of $220^{\prime} \times
160^{\prime}$ was surveyed down to a limiting surface brightness of 
$1\sigma = 26.5 R $ mag~arcsec$^{-2}$
within an aperture of 2 arcsec FWHM.

TT02 reported deeper observations of the 0.5 sq.~deg. core of the NGC~1407 
Group obtained with SuprimeCam on Subaru Telescope.  The current observations
of 20 times more area were designed to cover the entire virialized region
around NGC~1407.  In retrospect, we fail slightly to fully cover that region
and a small correction for incompletion will be described.

\subsection{Spectroscopy observations}

We undertook observations using the blue side of the Low Resolution
Imaging Spectrograph (LRIS: Oke et al.~1995) on the Keck~I 10m
telescope, using a 1\arcsec\ slit and a 600 lines~in$^{-1}$ grating.
Data were acquired on 27 November, 2003.  Observations were affected by
extensive cirrus and we obtained spectra for only four galaxies.  All
turned out to be group members.  One, N1407--041, also known as APMUK
B03381296-184859.2, is revealed to be a second case with a very large
blueshift ($-932$ km~s$^{-1}$ with respect to the group mean). 
It's location in close proximity to NGC~1407 is shown in Figure 3. 

\section{Sample}

Our ultimate intention is to get a listing of group members that is as
complete as possible.  An unambiguous way to establish membership is
to measure a velocity, but velocities can only be obtained for
the brighter galaxies.  We therefore need to rely on other information
to evaluate whether a galaxy lies in the group or
not.  
Dwarf galaxies have low surface brightnesses.  If
we see a low-surface-brightness galaxy in the direction of the group
of a type that we do not see in blank fields, there is a high probability that
this galaxy is a low-luminosity member of the group as opposed to a
high-luminosity background galaxy.  Other information, such as galaxy
morphology, can also be useful.  For example, if the
low-surface-brightness galaxy exhibits spiral structure, 
it is probably a luminous background galaxy
so there is a
low probability that it is a dwarf group member.  Morphology
assessments are subjective but their reliability can be tested.
For an extensive discussion of the issues
involved with the discovery and confirmation of dwarf galaxies, see 
Flint at al.~(2001).

\subsection{Establishing membership using velocity information}

A search of the NASA/IPAC Extragalactic Database (NED) reveals 29 galaxies
in our survey area have known velocities less than 3000~km s$^{-1}$.
Two additional galaxies have velocities within this limit in the second 
6 Degree Field (6dF) data release (Jones et al.~2005).  
Our Keck observations gave velocities for four more galaxies, 
all of which were in the group.

Consequently there are now a total of 35 group members with known velocities,
including almost all candidate members brighter than $M_R=-17$.  The rms
velocity dispersion for these galaxies is $387 \pm 65$~km s$^{-1}$.
The projected harmonic mean radius of all group candidates
is 446 kpc.  These quantities lead to a virial mass of $7.3 \times
10^{13} M_{\odot}$.  From this survey
we measure a total group $R$ band luminosity of 
$2.16 \times 10^{11} L_{\odot}$.
The mass to light ratio is then very large: $M/L_R = 340 M_{\odot}/L_{\odot}$.

\subsection{Establishing probability of membership using morphology information}

We are faced with the problem of establishing membership for galaxies that are too faint to be amenable to spectroscopy.
The uncertainties brought about by a purely
staristical background subtraction are large and that is why we cannot use this technique.
In the 10.05 square degree survey
area, we expect about 50 candidate members with $21>R>20$, based on knowledge of the field luminosity function (the 
uncertainty in this function corresponds to about a factor of two in this context).  
In this same 10.05 square degrees
of sky, we expect to find
about 20,000 background galaxies in this magnitude interval (see for
example Figure 6 of Trentham 1997).
The Poisson uncertainty in this number of background galaxies is about
140 but the primary source of uncertainty is large-scale structure
along the line of sight.  The precise uncertainty depends on exactly
where we are looking, but for an area of this size, we would expect a
field-to-field variance in the background counts of several times the
Poisson uncertainty, about several hundred galaxies.
It will therefore not be possible to identify members
at any satisfactory level of significance without some other criterion.

Our strategy has been to study such nearby groups that
    the commonly known dwarf galaxies would all be very well resolved
    and distinguishable from background galaxies based on surface
    brightness and morphological criteria.  Admittedly, ultracompact
    dwarfs would excape our attention but so far there is no indication
    that such high-surface-brightness objects have significant impact
    on counts at the faint end of the luminosity function.
 Initially in our program we had no proof that our methodology was
    valid.  However, the analysis of the NGC 5846 group gave us
    assurance that our selection criteria were working well.  We had
    complete spectroscopic coverage to $M_R=-15$ and sampled coverage to
    $M_R=-13.3$.  We have no proof that our selection criteria are working
    below this limit but fainter galaxies have even lower mean surface
    brightnesses and we feel that they separate very well from the
    background down to $M_R=-12$.  Below that limit the dimensions of the
    candidates get sufficiently small that confusion with the background
    becomes possible.

Following TT02 and MTT05,
we estimate that there is an appreciable
probability of membership if the inner (ICP) and outer (OCP)
concentration parameters obey the following criteria for bright
($R<20$) galaxies:
\begin{equation}
{\rm ICP} = R (4.4\,\,{\rm arcsec}) - R (2.2\,\,{\rm arcsec}) < 0.7
\end{equation}
and 
\begin{equation}
{\rm OCP} = R (12\,\,{\rm arcsec}) - R (6\,\,{\rm arcsec}) < 0.4.
\end{equation}
For faint ($R>20$) galaxies we think that there is an appreciable probability of membership if the galaxy consists of 
more than three seeing disks with surface brightness $\mu_R < 24.5$ mag arcsec$^{-2} $.  These criteria follow from 
the properties of nearby low-luminosity galaxies.
TT02 show how these selection rules favor inclusion of dwarfs and exclusion
of background galaxies.
All galaxies that meet the selection criteria were examined by eye.  
We lowered our assessment of the probability of membership if we could see 
any spiral structure or if the galaxy was very flat, these being properties normally found among high-luminosity galaxies.  
For each galaxy, we assigned a rating that describes our assessment of the membership probability (e.g .~TT02).
The ratings have the following meanings:
\vskip 1pt \noindent
``0'': membership confirmed from optical or HI spectroscopic data;
\vskip 1pt \noindent
``1'': probable member, but no spectroscopic or HI detection in the literature;
\vskip 1pt \noindent
``2'': possibly a member, but conceivably background;
\vskip 1pt \noindent
``3'': plausibly a member;
\vskip 1pt \noindent
``4'': passes ICP/OCP surface brightness test but almost certainly background given
the morphology of the galaxy and the properties of the background
fields studied at various stages during this project;
\vskip 1pt \noindent
``5'': fails ICP/OCP surface brightness test.
The vast majority (several thousand with $R<25$) of galaxies  
were rated 5.

The discussion below will point out that the vocabulary `possible' and 
`plausible' were too conservative.  Almost all `possible' candidates
turn out to be members and a substantial fraction of `plausible' candidates
are members.

The analysis of the NGC~5846 group by MTT05 has strengthened the basis of
the qualitative ratings.  That group has the fortune to lie in a part of
the sky with Sloan Digital Sky Survey (SDSS) spectroscopy, which provides
essentially complete velocity information to $M_R \sim -15$.  Our own Keck LRIS
observations provided scattered velocity information to $M_R \sim -13.4$.
As a result we could investigate membership issues with an extensive sample 
of 85 velocities.  The investigation was augmented by a correlation analysis
of the sort that will be described for the current data in the next
section.  The analysis carried out on the NGC~5846 Group is directly 
relevant to the current situation because the two groups are morphologically
similar, they are at almost the same distances, and they have been observed 
with the same instrument and analyzed with the same procedures.

The conclusions reached in MTT05 are that essentially all targets rated 1 and 
2 are true group members and that essentially all targets rated 4 are in the
background.  Concerning objects rated 3, MTT05
concluded that 70\% of those brighter than $M_R=-13.4$ (were they they to lie at the group distance)
are members, 50\% of early-type galaxies fainter than this limit are members, and
none of the late-type galaxies fainter than this limit are members.  
We will review these conclusions in the light of the new NGC~1407 Group
observations.  

Our recipe for identifying group members fails for
high surface brightness objects.  The extensive SDSS spectroscopic 
observations of the NGC~5846 region confirmed membership for a small number 
of high surface
brightness ellipticals in the regime $-20 < M_R < -17$, including a
system quite similar to the unusual Local Group member M32.  Interestingly, 
the three most extreme high surface brightness objects in that group
are all in close proximity to the dominant elliptical NGC~5846 which
suggests they may be tidally
truncated.  Spectroscopy, therefore, provides a useful complement to
our procedures, and a survey to a faint limit
around NGC~1407 would probably pick up a few currently unsuspected 
group members.
Compact objects have been identified through spectroscopy
in other dense groups (Phillipps et al. 2001) but they 
have not been
found in numbers that impact on the luminosity function.

On occasion we misidentify objects
as background galaxies on
morphological grounds.  An example is the case of N5846--256 =
CGCG~021-013 in MTT05.  Its velocity places it inside the group, but
the object looks like a background giant galaxy because it has a
substantial dust lane.  Another example is the nucleated dwarf
elliptical N1407--044 = LSBG~F549-036 in the current sample because
this galaxy is so flat that we thought it was a background disk
galaxies with a bulge.  Galaxies like this are extremely rare
but it is possible that a small number of the galaxies we rated ``4''
are actually members.

It is useful to make a comparison with the more sensitive study of the 
central part of the NGC~1407 Group reported by TT02, based on
Subaru 8m observations.  In Figure~4, we show the difference between the 
magnitudes derived from that survey 
and from the current study.  It is clear from this figure that the Subaru 
magnitudes are systematically brighter.  
Both studies measure the magnitude above the sky of each galaxy. 
The Subaru photometry is
deeper so the magnitudes represent total light within a fainter isophote.
For both datasets the median sky brightness was 20.6 mag arcsec$^{-2}$.
For the Subaru data the $1 \sigma$ noise within a 0.202 arcsec square pixel 
was about 26 mag arcsec$^{-2}$ but it was 
25 mag arcsec$^{-2}$ within a 0.206 arcsec square pixel for the shallower CFHT data.
In principle, we could fit exponential disk profiles to the CFHT images
and extrapolate to determine total magnitudes but uncertainties 
for these extremely low surface brightness galaxies would
be large (Tully et al.~1996).

\begin{figure}
\begin{center}
\vskip-4mm
\psfig{file=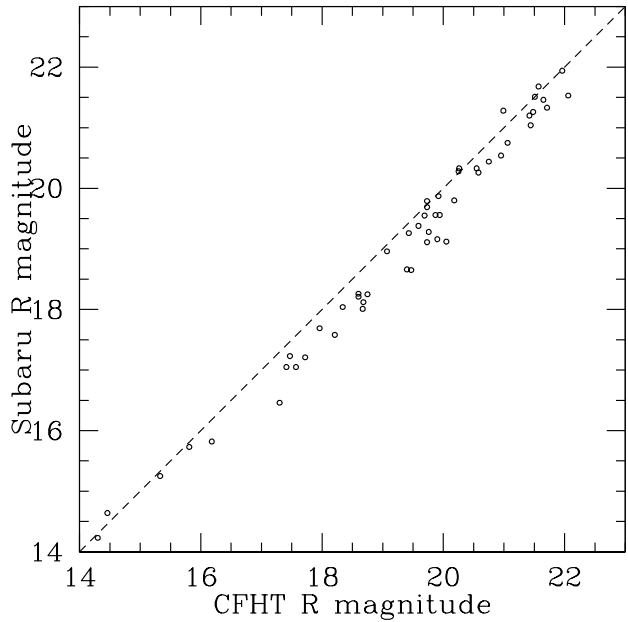, width=8.65cm}
\end{center}
\vskip-3mm
\caption{
The magnitudes of galaxy in the current survey compared to the magnitudes of the same galaxies measured by TT02
for all galaxies rated 0--3 in both studies.}
\end{figure}

%

The following few galaxies drew attention with this
comparison between the CFHT and Subaru data: 

\noindent
{\bf N1407-078 = TT02:15}:
This compact elliptical was rated 5 in the current data and 3 in the Subaru data, which is
surprising since most objects tend to be a bit more diffuse in the shallower CFHT data.

\noindent
{\bf N1407-125}:
This compact galaxy was rated 5 in the Subaru data, but it looks
more diffuse in the less deep CFHT data where it was rated 3.

\noindent
{\bf N1407-147 = TT02:26}: 
Once more, this compact elliptical was rated 5 in the current data and 3 in the Subaru data;
this is unusual for the reasons described above.
It was right on the edge of the chip in the Subaru data, which might explain why it was a 
bit fuzzy and was rated 3 there.

\noindent
{\bf N1407-160 = TT02:33}:
This faint galaxy was rated 5 in the current data and 3 in the Subaru data.
It is, however, marginal and at a stretch could be rated 3.

\noindent
{\bf N1407-164}:
This galaxy was rated 5 in deeper Subaru data but this rating was marginal and 
a rating of 3 is also acceptable.

\noindent
{\bf N1407-168}:
Like N1407-125, this compact galaxy was rated 5 in the Subaru data, but it looks
more diffuse in the less deep CFHT data where it was rated 3.

\noindent
{\bf N1407-187}:
This galaxy fell in a gap between CCD chips in the Subaru data.

\noindent
{\bf N1407-203}:
This compact galaxy was rated 5 in the Subaru data, but it was rated 3
in the shallower CFHT data.

\noindent
{\bf N1407-239}:
This galaxy fell in a gap between CCD chips in the Subaru data.

\noindent
{\bf N1407-218 = TT02:50}:
This galaxy was a rated 5 in the current survey because it looked like a double galaxy.  
The rating of 3 given by TT02 is marginal.

\noindent
{\bf N1407-241 = TT02:48}:
This galaxy was marginally detected in the shallower CFHT data.
It was indistinguishable from a large number of very faint 
low-surface-brightness objects in the data, most of which are 
flat fielding artifacts. 

\noindent
{\bf N1407-245 = TT02:53}:
Once more, this galaxy was marginally detected in the shallower CFHT data
but was indistinguishable from the large number of faint low-surface-brightness objects in the data.

\noindent
{\bf N1407-254}:
This compact galaxy was rated 5 in the Subaru data, but it was rated 3
in the shallower CFHT data.

\noindent
{\bf N1407-261 = TT02:57}:
Yet again this galaxy was marginally detected in the shallower CFHT data
but was indistinguishable from the large number of faint low-surface-brightness objects in the data.

\noindent
{\bf N1407-263}:
This galaxy resolved into smaller galaxies in the deeper Subaru data and was rated 4.

\noindent
{\bf N1407-265}:
This very faint galaxy looked like it resolved into two galaxies in the Subaru data and was rated 4.

\noindent
{\bf N1407-267 = TT02:55}:
This galaxy was undetected in the shallower CFHT data.

\noindent
{\bf N1407-268 = TT02:58}:
This galaxy was undetected in the shallower CFHT data.

The Subaru observations allowed us to discover fainter candidates but only in
a limited area.  The objects identified uniquely with Subaru are included
in Table 2 but are flagged and not included in the statistically complete
sample.

\begin{table*}
\caption{The NGC 1407 Group Sample}
{\vskip 0.55mm} {$$\vbox{ \halign {\hfil #\hfil && \quad \hfil #\hfil \cr
\noalign{\hrule \medskip}
ID  & Name &  Type & Rating & $V_h$ & $\alpha$ (J2000) & $\delta$ (J2000) & $R_{300}$ & $R$ & $M_R$ & ID$_{\rm TT}$ & Rat$_{\rm TT}$ & $R_{\rm TT}$
                      &\cr
    &            &      &    &  km/s    &  &             &        &        &         &    &   &          &\cr
\noalign{\smallskip \hrule \smallskip}
\cr
  1 & NGC 1407   & E    &  0 & 1779 & 3 40 11.9 & -18 34 49 & 13.78 &   9.01 & -23.16 &  1 & 0 &    10.23 &\cr
  2 & NGC 1400   & S0   &  0 &  558 & 3 39 30.8 & -18 41 17 & 13.21 &  10.37 & -21.80 &  2 & 0 &    10.57 &\cr
  3 & NGC 1440   & S0   &  0 & 1534 & 3 45 02.9 & -18 15 57 & 12.98 &  10.81 & -21.47 &    &   &    &\cr
  4 & NGC 1452   & Sa   &  0 & 1737 & 3 45 22.3 & -18 38 01 & 13.51 &  11.35 & -20.90 &    &   &    &\cr
  5 & NGC 1393   & S0   &  0 & 2185 & 3 38 38.6 & -18 25 40 & 12.90 &  11.74 & -20.40 &  3 & 0 &    11.45 &\cr
  6 & NGC 1383   & S0   &  0 & 1948 & 3 37 39.2 & -18 20 22 & 14.09 &  11.84 & -20.36 &    &   &    &\cr
  7 & IC 346     & S0   &  0 & 2013 & 3 41 44.7 & -18 16 01 & 14.99 &  12.24 & -19.94 &    &   &    &\cr
  8 & NGC 1359         & Sm   &  0 & 1966 & 3 33 47.7 & -19 29 31 & 16.30 &  12.16 & -19.93 &    &   &    &\cr
  9 & ESO 548-G047 & S0   &  0 & 1606 & 3 34 43.5 & -19 01 44 & 14.83 &  12.23 & -19.89 &    &   &    &\cr
 10 & ESO 548-G068 & S0   &  0 & 1693 & 3 40 19.2 & -18 55 53 & 15.12 &  12.75 & -19.44 &    &   &    &\cr
 11 & IC 343       & S0   &  0 & 1841 & 3 40 07.1 & -18 26 37 & 15.16 &  12.79 & -19.39 &  4 & 0 &    12.87 &\cr
 12 & ESO 548-G044 & Sa   &  0 & 1696 & 3 34 19.2 & -19 25 28 & 15.38 &  12.86 & -19.23 &    &   &    &\cr
 13 & ESO 548-G079 & dE,N &  0 & 2016 & 3 41 56.1 & -18 53 43 & 15.62 &  13.03 & -19.18 &    &   &    &\cr
 14 & ESO 548-G033 & S0pec&  0 & 1699 & 3 32 28.6 & -18 56 55 & 15.32 &  13.04 & -19.05 &    &   &    &\cr
 15 & ESO 548-G064 & S0   &  0 & 1694 & 3 40 00.1 & -19 25 35 & 15.19 &  13.44 & -18.75 &    &   &    &\cr      
 16 & IC 345       & Sa   &  0 & 1335 & 3 41 09.1 & -18 18 51 & 14.30 &  13.43 & -18.75 &    &   &    &\cr
 17 & ESO 548-G076 & S0/a &  0 & 1471 & 3 41 31.8 & -19 54 19 & 16.90 &  13.48 & -18.70 &    &   &    &\cr      
 18 & NGC 1390     & Sa   &  0 & 1207 & 3 37 52.2 & -19 00 30 & 16.35 &  13.64 & -18.52 &    &   &    &\cr      
 19 & ESO 548-G063 & Sdm  &  0 & 1988 & 3 39 34.8 & -20 00 53 & 16.95 &  14.03 & -18.12 &    &   &    &\cr      
 20 & ESO 548-G082 & Sd   &  0 & 1716 & 3 42 43.3 & -17 30 26 & 18.23 &  14.39 & -17.90 &    &   &    &\cr      
 21 & LEDA 074868  & E    &  0 & 1575 & 3 40 15.9 & -19 04 54 & 16.56 &  14.32 & -17.85 &    &   &    &\cr      
 22 & ESO 548-G043 & Sd   &  0 & 1931 & 3 34 10.5 & -19 33 30 & 15.64 &  14.27 & -17.81 &    &   &    &\cr      
 23 & ESO 548-G002 & Sm   &  0 & 1111 & 3 42 57.3 & -19 01 12 & 17.70 &  14.39 & -17.80 &    &   &    &\cr      
 24 & ESO 548-G065 & Sa   &  0 & 1221 & 3 40 02.7 & -19 22 00 & 17.82 &  14.40 & -17.80 &    &   &    &\cr
 25 & LEDA 074886  & E    &  0 & 1308 & 3 40 43.2 & -18 38 43 & 16.29 &  14.30 & -17.79 &  5 & 0 &    14.23 &\cr
 26 & APMBGC 549+118-079 & Sd   &  0 & 1979 & 3 44 02.5 & -18 28 18 & 17.80 &  14.45 & -17.75 &    &   &    &\cr
 27 & ESO 548-G073  & Sdm &  0 & 989  & 3 41 04.4 & -19 05 40 & 17.50 &  14.46 & -17.74 &    &   &    &\cr
 28 & ESO 549-G007  & E   &  0 & 1527 & 3 44 11.5 & -19 19 13 & 17.11 &  14.43 & -17.72 &    &   &    &\cr
 29 & APMBGC 548-110-078 & E    &  0 & 1595 & 3 40 52.7 & -18 28 40 & 16.88 &  14.46 & -17.71 &  6 & 0 &    14.64 &\cr
 30 & LEDA 135109   & E   &  3 &      & 3 43 42.7 & -19 49 48 & 15.74 &  14.50 & -17.66 &    &   &    &\cr
 31 & LEDA 074924   & E   &  2 &      & 3 41 59.9 & -18 42 47 & 16.85 &  14.76 & -17.42 &    &   &    &\cr
 32 & APMBGC 548-118-089 & E    &  2 &      & 3 41 29.8 & -18 15 50 & 17.23 &  14.87 & -17.32 &    &   &    &\cr
 33 & ESO 548-G072  & Sm  &  0 & 2034 & 3 41 00.3 & -19 27 19 & 18.34 &  15.01 & -17.20 &    &   &    &\cr
 34 & LSBG F549-032 & dE,N&  2 &      & 3 42 47.5 & -17 34 04 & 17.99 &  15.30 & -17.00 &    &   &    &\cr
 35 & APMBGC 548-122-018 & Epec &  0 &  1913 & 3 41 49.8 & -19 34 53 & 17.98 &  15.32 & -16.87 &    &   &    &\cr
 36 & LEDA 074838   & dE,N&  2 &      & 3 39 23.1 & -18 45 30 & 17.67 &  15.33 & -16.84 &  7 & 1 &    15.25 &\cr
 37 & LSBG F548-012  & dE,N &  2 &      & 3 38 42.1 & -18 53 59 & 18.54 &  15.52 & -16.66 &    &   &    &\cr
 38 & 2MASXi J0341186-180206 & E    &  3 &      & 3 41 18.6 & -18 02 07 & 17.36 &  15.76 & -16.64 &    &   &    &\cr
 39 & APMUKS B033659.29-185352.7 & dE,N &  2 &      & 3 39 14.5 & -18 44 10 & 17.57 &  15.81 & -16.35 &  8 & 2 &    15.73 &\cr
 40 & 2MASXi J0341237-183808 & E    &  3 &      & 3 41 23.7 & -18 38 08 & 17.76 &  16.05 & -16.12 &    &   &    &\cr
 41 & LSBG F548-006 & dE,N &  0 &  698 & 3 40 33.8 & -18 39 03 & 18.66 &  16.18 & -16.00 &  9 & 1 &    15.82 &\cr
 42 & LSBG F548-016 & dE,N &  2 &      & 3 35 10.7 & -18 55 51 & 18.59 &  16.19 & -15.93 &    &   &    &\cr
 43 & LSBG F549-023 & dE,N &  2 &      & 3 44 10.7 & -17 49 59 & 18.69 &  16.40 & -15.90 &    &   &    &\cr
 44 & LSBG F549-036  & dE,N &  0 & 2101 & 3 41 46.6 & -17 44 22 & 18.44 &  16.45 & -15.75 &    &   &    &\cr
 45 & 2MASX J03425744-1900411 & E    &  2 &      & 3 42 57.4 & -19 00 41 & 17.32 &  16.47 & -15.73 &    &   &    &\cr
 46 & N1400GR:[FS90] 079  & dE   &  3 &      & 3 41 10.1 & -17 38 13 & 18.79 &  16.60 & -15.67 &    &   &    &\cr
 47 & LSBG F549-038   & dE,N &  0 & 1893 & 3 40 49.7 & -18 50 50 & 18.59 &  16.70 & -15.52 &    &   &    &\cr
 48 & LSBG F548-026   & dE,N &  0 & 1492 & 3 39 33.9 & -19 15 58 & 18.49 &  16.67 & -15.51 &    &   &    &\cr
 49 & N1400GR:[FS90] 067 & dE,N &  3 &      & 3 40 47.4 & -18 37 11 & 18.73 &  16.71 & -15.46 &    &   &    &\cr
 50 & APMUKS B033801.24-191219.0 & dI   &  3 &      & 3 40 16.0 & -19 02 41 & 18.75 &  16.85 & -15.33 &    &   &    &\cr
 51 & APMUKS B033224.45-192557.3 & dE   &  2 &      & 3 34 39.4 & -19 15 58 & 19.18 &  16.78 & -15.31 &    &   &    &\cr
 52 & LSBG F548-003 & dI   &  3 &      & 3 42 54.8 & -17 33 23 & 18.92 &  17.11 & -15.20 &    &   &    &\cr
 53 & LSBG F548-032 & dE   &  2 &      & 3 33 11.5 & -19 19 02 & 18.90 &  17.10 & -14.99 &    &   &    &\cr
 54 & LSBG F549-021  & dE,N &  3 &      & 3 45 09.4 & -17 55 08 & 19.11 &  17.27 & -14.99 &    &   &    &\cr
 55 & LEDA 074762  & dE   &  1 &      & 3 36 05.9 & -17 52 36 & 20.12 &  17.33 & -14.88 &    &   &    &\cr
\cr
\noalign{\smallskip \hrule}
\noalign{\smallskip}\cr}}$$}
\end{table*}

\begin{table*}
{\vskip 0.55mm} {$$\vbox{ \halign {\hfil #\hfil && \quad \hfil #\hfil \cr
\noalign{\hrule \medskip}
ID  & Name &  Type & Rating & $V_h$ & $\alpha$ (J2000) & $\delta$ (J2000) & $R_{300}$ & $R$ & $M_R$ & ID$_{\rm TT}$ & Rat$_{\rm TT}$ & $R_{\rm TT}$
                      &\cr
    &            &      &    &  km/s    &  &             &        &        &         &    &   &          &\cr
\noalign{\smallskip \hrule \smallskip}
\cr
 56 & LEDA 074845  & dI   &  1 &      & 3 39 42.0 & -18 40 03 & 20.51 &  17.30 & -14.87 & 10 & 1 &    16.46 &\cr
 57 & LEDA 074830 & dE,N &  1 &      & 3 39 04.5 & -18 31 56 & 19.29 &  17.41 & -14.75 & 12 & 2 &    17.05 &\cr
 58 & LSBG F548-011 & dE,N &  2 &      & 3 39 04.6 & -18 21 36 & 19.17 &  17.47 & -14.70 & 14 & 3 &    17.23 &\cr
 59 & & dE   & 2 &      & 3 37 51.3 & -18 38 45 & 19.41 &  17.57 & -14.63 &    &   &    &\cr
 60 & APMUKS B033247.49-182749.0 & dE   &  3 &      & 3 35 03.4 & -18 17 52 & 19.34 &  17.56 & -14.63 &    &   &    &\cr
 61 & LSBG F549-037 & dE   &  3 &      & 3 40 54.8 & -17 31 30 & 19.88 &  17.64 & -14.63 &    &   &    &\cr
 62 & LEDA 074847 & dE,N &  1 &      & 3 39 45.4 & -18 30 16 & 20.13 &  17.57 & -14.60 & 11 & 1 &    17.05 &\cr
 63 &  & dI   &  3 &      & 3 34 32.8 & -19 01 03 & 19.07 &  17.54 & -14.58 &    &   &    &\cr
 64 & APMUKS B033659.86-195850.8 & dE   &  3 &      & 3 39 13.8 & -19 49 08 & 19.73 &  17.63 & -14.54 &    &   &    &\cr
 65 & N1400GR:[FS90] 058 & dE,N &  1 &      & 3 40 28.2  & -18 39 22 & 19.37 &  17.72 & -14.46 & 13 & 1 &    17.21 &\cr
 66 & APMUKS B033423.76-194449.4  & dI   &  1 &      & 3 36 38.2 & -19 34 57 & 19.53 &  17.76 & -14.34 &    &   &    &\cr
 67 & LEDA 074891 & dE,N &  1 &      & 3 40 50.0 & -18 05 29 & 20.05 &  17.90 & -14.29 &    &   &    &\cr
 68 & APMUKS B034054.06-195241.1   & dI/E &  1 &      & 3 43 08.0 & -19 43 13 & 21.07 &  17.92 & -14.27 &    &   &    &\cr
 69 & LSBG F548-005 & dE,N &  3 &      & 3 42 13.2 & -18 56 51 & 19.57 &  17.96 & -14.24 &    &   &    &\cr
 70 &  & dE/N &  1 &      & 3 40 00.9 & -18 48 14 & 19.70 &  17.96 & -14.21 & 16 & 1 &    17.69 &\cr
 71 & APMUKS B033739.33-185912.5 & dE   &  3 &      & 3 39 54.4 & -18 49 33 & 19.52 &  18.04 & -14.13 &    &   &    &\cr
 72 & N1400GR:[FS90] 012 & dE   &  1 &      & 3 37 52.2 & -18 55 27 & 20.48 &  18.05 & -14.12 &    &   &    &\cr
 73 & APMUKS B033150.50-193739.6 & dE   &  3 &      & 3 34 05.0 & -19 27 43 & 19.96 &  18.03 & -14.06 &    &   &    &\cr
 74 & LSBG F549-025 & dE   &  3 &      & 3 43 54.4 & -18 30 07 & 19.89 &  18.21 & -13.99 &    &   &    &\cr
 75 & & dE   &  1 &      & 3 38 26.2 & -18 00 07 & 20.76 &  18.28 & -13.94 &    &   &    &\cr
 76 & N1400GR:[FS90] 059 & dE,N &  1 &      & 3 40 28.8 & -18 58 17 & 20.68 &  18.29 & -13.89 &    &   &    &\cr
 77 & APMUKS B033150.50-193739.6 & dE   &  3 &      & 3 34 05.2 & -19 27 37 & 20.33 &  18.22 & -13.87 &    &   &    &\cr
 78 & APMUKS B033912.64-183429.9 & dE,N &  3*&      & 3 41 27.7 & -18 24 57 & 19.66 &  18.21 & -13.87 & 15 & 3 &    17.58 &\cr
 79 & N1400GR:[FS90] 032 & dE,N &  1 &      & 3 39 09.2 & -18 26 44 & 20.31 &  18.31 & -13.85 &    &   &    &\cr
 80 & N1400GR:[FS90] 033 & dE,N &  2 &      & 3 39 09.7 & -18 37 28 & 19.90 &  18.34 & -13.82 & 18 & 2 &    18.04 &\cr
 81 & APMUKS B033329.47-184051.6 & dE   &  2 &      & 3 35 45.1 & -18 30 58 & 19.96 &  18.34 & -13.80 &    &   &    &\cr
 82 &         & dE,N &  3 &      & 3 41 54.0 & -17 38 50 & 20.23 &  18.53 & -13.76 &    &   &    &\cr
 83 & LSBG F549-018        & dI   &  3 &      & 3 45 48.4 & -18 50 36 & 20.41 &  18.48 & -13.75 &    &   &    &\cr
 84 & LEDA 074784        & dE   &  3 &      & 3 37 03.7 & -18 58 18 & 19.66 &  18.43 & -13.71 &    &   &    &\cr
 85 &   & dI   &  3 &      & 3 39 19.3 & -19 19 40 & 21.19 &  18.54 & -13.62 &    &   &    &\cr
 86 &  & VLSB &  1 &      & 3 38 49.1 & -18 42 13 & 21.48 &  18.60 & -13.58 & 22 & 1 &    18.26 &\cr
 87 & APMUKS B033057.83-175450.4 & dI    &  3 &     & 3 33 14.3 & -17 44 50 & 20.46 &  18.63 & -13.58 &    &   &    &\cr
 88 & LEDA 074854         & dE,N &  2 &      & 3 39 53.3 & -18 37 17 & 19.93 &  18.60 & -13.57 & 20 & 2 &    18.21 &\cr
 89 & APMUKS B033402.22-194037.3      & dE,N &  1 &      & 3 36 16.8 & -19 30 43 & 19.86 &  18.53 & -13.57 &    &   &    &\cr
 90 & APMUKS B034324.69-182212.5 & dE,N &  1 &      & 3 45 40.1 & -18 12 55 & 20.90 &  18.72 & -13.56 &    &   &    &\cr
 91 & APMUKS B034020.33-195428.2  & dE   &  3 &      & 3 42 34.2 & -19 44 58 & 20.23 &  18.63 & -13.54 &    &   &    &\cr
 92 & N1400GR:[FS90] 044 & dE,N &  2 &      & 3 39 48.8 & -18 53 20 & 20.23 &  18.65 & -13.52 &    &   &    &\cr
 93 & LEDA 074814  & dE   &  1 &      & 3 38 28.3 & -18 00 05 & 20.65 &  18.71 & -13.51 &    &   &    &\cr
 94 & LEDA 074804  & dE,N &  2 &      & 3 37 58.8 & -18 42 42 & 20.02 &  18.70 & -13.50 &    &   &    &\cr
 95 & & dE   &  3 &      & 3 36 05.7 & -18 23 02 & 20.03 &  18.67 & -13.50 &    &   &    &\cr
 96 & LEDA 074857  & dE,N &  1 &      & 3 39 59.5 & -18 29 20 & 20.59 &  18.68 & -13.50 & 19 & 1 &    18.12 &\cr
 97 & N1400GR:[FS90] 032  & dE,N &  1 &      & 3 39 09.1 & -18 26 42 & 20.19 &  18.67 & -13.49 & 17 & 2 &    18.01 &\cr
 98 &         & dE,N &  1 &      & 3 40 59.0 & -17 43 04 & 22.25 &  18.75 & -13.49 &    &   &    &\cr
 99 & N1400GR:[FS90] 102 & dE   &  1 &      & 3 43 25.2 & -17 51 16 & 20.83 &  18.86 & -13.47 &    &   &    &\cr
100 & N1400GR:[FS90] 045 & dE   &  2 &      & 3 39 51.3 & -18 27 59 & 20.31 &  18.75 & -13.43 & 21 & 1 &    18.25 &\cr
101 & APMUKS B034300.92-191231.4  & dE,N &  2 &      & 3 45 15.4 & -19 03 11 & 21.38 &  18.81 & -13.41 &    &   &    &\cr
102 &         & dE   &  1 &      & 3 40 07.4 & -17 58 49 & 21.09 &  18.79 & -13.40 &    &   &    &\cr
103 & APMUKS B033515.30-200435.9  & dE   &  3 &      & 3 37 29.4 & -19 54 49 & 20.28 &  18.71 & -13.39 &    &   &    &\cr
104 &         & dE   &  1 &      & 3 39 32.4 & -19 25 16 & 21.01 &  18.80 & -13.37 &    &   &    &\cr
105 &         & VLSB &  1 &      & 3 37 28.7 & -19 11 15 & 22.25 &  18.77 & -13.37 &    &   &    &\cr
106 & APMUKS B033220.92-201714.6 & dE/I &  2 &      & 3 34 34.9 & -20 07 16 & 20.79 &  18.74 & -13.33 &    &   &    &\cr
107 &         & dE/I &  2 &      & 3 46 50.0 & -19 48 28 & 21.30 &  18.96 & -13.32 &    &   &    &\cr
108 &         & dI   &  3 &      & 3 39 27.4 & -17 30 35 & 20.38 &  18.92 & -13.32 &    &   &    &\cr
109 &         & dE   &  1 &      & 3 35 40.9 & -18 43 41 & 20.87 &  18.85 & -13.28 &    &   &    &\cr
110 & LEDA 074765 & dE   &  3 &      & 3 36 07.2 & -18 22 43 & 20.00 &  18.92 & -13.25 &    &   &    &\cr
\cr
\noalign{\smallskip \hrule}
\noalign{\smallskip}\cr}}$$}
\end{table*}

\begin{table*}
{\vskip 0.55mm} {$$\vbox{ \halign {\hfil #\hfil && \quad \hfil #\hfil \cr
\noalign{\hrule \medskip}
ID  & Name &  Type & Rating & $V_h$ & $\alpha$ (J2000) & $\delta$ (J2000) & $R_{300}$ & $R$ & $M_R$ & ID$_{\rm TT}$ & Rat$_{\rm TT}$ & $R_{\rm TT}$
                      &\cr
    &            &      &    &  km/s    &  &             &        &        &         &    &   &          &\cr
\noalign{\smallskip \hrule \smallskip}
\cr
111 &         & dE   &  2 &      & 3 39 33.0 & -18 07 16 & 20.65 &  18.97 & -13.23 &    &   &    &\cr
112 & APMUKS B033605.04-182859.2 & dI   &  3 &      & 3 38 20.8 & -18 19 15 & 19.98 &  19.00 & -13.17 &    &   &    &\cr
113 & LEDA 074878 & dE,N &  1 &      & 3 40 30.7 & -17 46 00 & 20.33 &  19.06 & -13.16 &    &   &    &\cr
114 & N1400GR:[FS90] 031 & dE,N &  3 &      & 3 39 07.0 & -18 59 16 & 20.19 &  19.04 & -13.13 &    &   &    &\cr
115 &         & dE   &  1 &      & 3 39 51.0 & -18 32 23 & 21.75 &  19.07 & -13.10 & 25 & 1 &    18.96 &\cr
116 &         & dE/I &  2 &      & 3 35 24.0 & -19 41 40 & 20.80 &  19.03 & -13.06 &    &   &    &\cr
117 &         & dE   &  1 &      & 3 37 39.0 & -19 09 50 & 21.05 &  19.11 & -13.04 &    &   &    &\cr
118 & APMUKS B034533.53-193604.0 & dE/I &  2 &      & 3 47 47.8 & -19 26 54 & 21.12 &  19.22 & -13.00 &    &   &    &\cr
119 & & dE,N &  2 &      & 3 35 26.0 & -19 41 27 & 20.81 &  19.10 & -12.99 &    &   &    &\cr
120 & & dE,N &  1 &      & 3 40 28.1 & -18 19 31 & 21.17 &  19.19 & -12.98 &    &   &    &\cr
121 & N1400GR:[FS90] 007 & dE   &  3 &      & 3 37 17.0 & -18 10 31 & 20.39 &  19.23 & -12.94 &    &   &    &\cr
122 & & dE   &  1 &      & 3 35 09.9 & -18 38 33 & 21.71 &  19.23 & -12.93 &    &   &    &\cr
123 & & dE   &  2 &      & 3 35 25.2 & -18 41 55 & 20.56 &  19.23 & -12.92 &    &   &    &\cr
124 & & dE   &  1 &      & 3 40 55.8 & -18 49 15 & 21.32 &  19.35 & -12.87 &    &   &    &\cr
125 & N1400GR:[FS90] 046 & dE,N &  3 &      & 3 39 52.6 & -18 46 51 & 20.11 &  19.32 & -12.85 &    & 4 &    &\cr
126 & & dE   &  1 &      & 3 40 33.7 & -19 07 07 & 22.41 &  19.35 & -12.83 &    &   &    &\cr
127 & & dE/I &  2 &      & 3 36 43.0 & -18 02 11 & 20.53 &  19.37 & -12.81 &    &   &    &\cr
128 & & dE,N &  1 &      & 3 38 11.5 & -18 22 56 & 21.10 &  19.40 & -12.78 & 24 & 1 &    18.66 &\cr
129 & & dE   &  2 &      & 3 44 50.1 & -19 18 31 & 20.87 &  19.40 & -12.78 &    &   &    &\cr
130 & & dE   &  2 &      & 3 42 02.3 & -18 26 37 & 21.20 &  19.43 & -12.76 & 29 & 2 &    19.26 &\cr
131 & & dE   &  3 &      & 3 45 04.4 & -18 36 18 & 20.63 &  19.50 & -12.74 &    &   &    &\cr
132 & & dE,N &  1 &      & 3 40 02.6 & -18 22 36 & 21.15 &  19.45 & -12.73 &    &   &    &\cr
133 & & dE   &  3 &      & 3 41 44.3 & -19 11 35 & 20.46 &  19.46 & -12.73 &    &   &    &\cr
134 & & dE,N &  1 &      & 3 40 02.6 & -18 22 36 & 20.88 &  19.47 & -12.71 & 23 & 1 &    18.65 &\cr
135 & & dE   &  3 &      & 3 35 42.2 & -18 03 19 & 20.67 &  19.51 & -12.68 &    &   &    &\cr
136 & N1400GR:[FS90] 014 & dE   &  1 &      & 3 37 54.58 & -19 00 18 & 20.26 &  19.50 & -12.66 &    &   &    &\cr
137 & & dE   &  2 &      & 3 35 55.2 & -18 41 29 & 20.54 &  19.48 & -12.65 &    &   &    &\cr
138 & & VLSB &  2 &      & 3 39 46.1 & -18 13 06 & 23.41 &  19.55 & -12.64 &    &   &    &\cr
139 & & dI/V &  2 &      & 3 41 31.1 & -18 52 35 & 23.31 &  19.61 & -12.61 &    &   &    &\cr
140 & N1400GR:[FS90] 063 & dE,N &  2 &      & 3 40 44.0 & -18 44 40 & 20.76 &  19.59 & -12.61 & 31 & 2 &    19.38 &\cr
141 & & dE,N &  3 &      & 3 39 00.0 & -18 06 30 & 20.78 &  19.62 & -12.59 &    &   &    &\cr
142 & & dE   &  3 &      & 3 37 03.2 & -18 48 58 & 20.87 &  19.60 & -12.54 &    &   &    &\cr
143 & N1400GR:[FS90] 095 & dE   &  2 &      & 3 42 31.6 & -18 32 29 & 21.38 &  19.66 & -12.51 &    &   &    &\cr
144 & & dE   &  2 &      & 3 35 38.4 & -18 54 52 & 20.46 &  19.64 & -12.50 &    &   &    &\cr
145 & & dI   &  2 &      & 3 39 50.8 & -18 22 51 & 21.43 &  19.69 & -12.50 & 32 & 3 &    19.55 &\cr
146 & & dE   &  2 &      & 3 42 06.1 & -19 35 55 & 21.12 &  19.71 & -12.48 &    &   &    &\cr
147 & & dE   &  3*&      & 3 39 55.1 & -18 21 22 & 20.71 &  19.73 & -12.46 & 26 & 3 &    19.11 &\cr
148 & N1400GR:[FS90] 073 & dE,N &  1 &      & 3 40 56.6 & -18 39 22 & 21.50 &  19.73 & -12.45 & 36 & 2 &    19.79 &\cr
149 & & dE,N &  1 &      & 3 38 28.5 & -18 46 22 & 21.29 &  19.76 & -12.44 & 30 & 2 &    19.28 &\cr
150 & & dE,N &  2 &      & 3 39 42.0 & -18 42 58 & 21.42 &  19.73 & -12.43 & 35 & 2 &    19.69 &\cr
151 & N1400GR:[FS90] 115 & dE,N &  1 &      & 3 45 23.3 & -18 07 24 & 20.95 &  19.82 & -12.43 &    &   &    &\cr
152 & & dE   &  2 &      & 3 36 07.5 & -18 11 04 & 20.79 &  19.76 & -12.43 &    &   &    &\cr
153 & & dE   &  2 &      & 3 36 03.1 & -18 48 59 & 20.97 &  19.71 & -12.42 &    &   &    &\cr
154 & & dE/I &  3 &      & 3 35 15.9 & -19 18 26 & 20.98 &  19.69 & -12.41 &    &   &    &\cr
155 & & dE   &  3 &      & 3 39 25.9 & -18 52 05 & 20.70 &  19.77 & -12.40 &    &   &    &\cr
156 & & dE   &  2 &      & 3 41 42.5 & -18 59 11 & 21.36 &  19.84 & -12.36 &    &   &    &\cr
157 & & dE   &  1 &      & 3 37 38.6 & -18 21 37 & 20.89 &  19.83 & -12.36 &    &   &    &\cr
158 & & dE   &  2 &      & 3 37 35.0 & -18 33 27 & 20.99 &  19.84 & -12.35 &    &   &    &\cr
159 & & dE/I &  2 &      & 3 39 04.0 & -18 19 22 & 21.47 &  19.84 & -12.34 &    &   &    &\cr
160 & & dE   &  3*&      & 3 41 27.8 & -18 42 28 & 20.95 &  19.87 & -12.32 & 33 & 3 &    19.56 &\cr
161 & N1400GR:[FS90] 062 & dE,N &  1 &      & 3 40 37.6 & -18 32 45 & 21.09 &  19.90 & -12.27 & 28 & 2 &    19.16 &\cr
162 & & dE   &  2 &      & 3 33 19.9 & -18 41 09 & 21.53 &  19.84 & -12.26 &    &   &    &\cr
163 & & dE   &  2 &      & 3 42 17.8 & -18 47 19 & 21.32 &  19.94 & -12.26 &    &   &    &\cr
164 & & dE,N &  3 &      & 3 39 58.1 & -18 37 36 & 21.69 &  19.92 & -12.25 &    & 4 &    &\cr
165 & N1400GR:[FS90] 080 & dE   &  3 &      & 3 41 14.2 & -18 38 26 & 20.87 &  19.92 & -12.24 & 38 & 2 &    19.87 &\cr
\cr
\noalign{\smallskip \hrule}
\noalign{\smallskip}\cr}}$$}
\end{table*}

\begin{table*}
{\vskip 0.55mm} {$$\vbox{ \halign {\hfil #\hfil && \quad \hfil #\hfil \cr
\noalign{\hrule \medskip}
ID  & Name &  Type & Rating & $V_h$ & $\alpha$ (J2000) & $\delta$ (J2000) & $R_{300}$ & $R$ & $M_R$ & ID$_{\rm TT}$ & Rat$_{\rm TT}$ & $R_{\rm TT}$
                      &\cr
    &            &      &    &  km/s    &  &             &        &        &         &    &   &          &\cr
\noalign{\smallskip \hrule \smallskip}
\cr
166 & N1400GR:[FS90] 070 & dE   &  1 &      & 3 40 51.5 & -18 29 44 & 21.42 &  19.94 & -12.22 & 34 & 1 &    19.56 &\cr
167 & & dE,N &  1 &      & 3 39 13.9 & -18 53 43 & 21.43 &  19.96 & -12.22 &    &   &    &\cr
168 & & dE   &  3 &      & 3 38 11.5 & -18 24 20 & 20.68 &  20.04 & -12.15 &    & 4 &    &\cr
169 & & dE   &  3 &      & 3 38 10.6 & -18 18 58 & 21.72 &  20.03 & -12.14 &    &   &    &\cr
170 & & dE   &  2 &      & 3 35 31.5 & -19 50 06 & 21.45 &  19.97 & -12.12 &    &   &    &\cr
171 & & dI   &  2 &      & 3 33 41.2 & -19 00 45 & 21.34 &  19.99 & -12.11 &    &   &    &\cr
172 & & dE   &  2 &      & 3 35 06.4 & -17 38 08 & 21.31 &  20.10 & -12.10 &    &   &    &\cr
173 & & dE/I &  1 &      & 3 46 02.8 & -18 19 18 & 22.41 &  20.21 & -12.06 &    &   &    &\cr
174 & & dE   &  3 &      & 3 39 02.6 & -18 00 30 & 21.18 &  20.16 & -12.06 &    &   &    &\cr
175 & & dE,N &  1 &      & 3 38 52.0 & -18 26 00 & 21.19 &  20.05 & -12.03 & 27 & 1 &    19.12 &\cr
176 & & dE,N &  2 &      & 3 39 11.4 & -18 55 32 & 21.44 &  20.18 & -11.99 &    &   &    &\cr      
177 & & dE,N &  2 &      & 3 39 42.4 & -19 38 28 & 21.31 &  20.19 & -11.98 &    &   &    &\cr
178 & & dE,N &  2 &      & 3 39 22.3 & -18 31 59 & 21.53 &  20.18 & -11.98 & 37 & 1 &    19.80 &\cr
179 & & dE   &  3 &      & 3 40 26.4 & -20 06 50 & 21.17 &  20.03 & -11.96 &    &   &    &\cr      
180 & & dE   &  2 &      & 3 38 10.4 & -18 34 22 & 21.37 &  20.25 & -11.94 & 39 & 2 &    20.28 &\cr
181 & & dE,N &  2 &      & 3 43 01.6 & -18 37 35 & 21.39 &  20.23 & -11.94 &    &   &    &\cr
182 & & dE/I &  2 &      & 3 47 31.8 & -18 50 34 & 21.38 &  20.30 & -11.93 &    &   &    &\cr
183 & & dE,N &  2 &      & 3 38 38.8 & -18 42 16 & 21.01 &  20.26 & -11.93 & 41 & 3 &    20.33 &\cr
184 & & dE/I &  2 &      & 3 38 26.9 & -18 56 26 & 21.23 &  20.26 & -11.92 &    &   &    &\cr
185 & & dE,N &  2 &      & 3 40 31.2 & -17 54 05 & 21.63 &  20.29 & -11.91 &    &   &    &\cr
186 & & dI   &  2 &      & 3 42 32.4 & -19 21 56 & 22.08 &  20.26 & -11.90 &    &   &    &\cr      
187 & & dE,N &  3 &      & 3 38 33.1 & -18 44 08 & 21.32 &  20.30 & -11.90 &    &   &    &\cr
188 & & dE/I &  1 &      & 3 34 17.3 & -19 04 30 & 22.54 &  20.23 & -11.88 &    &   &    &\cr
189 & & dE   &  2 &      & 3 40 31.5 & -18 15 05 & 21.76 &  20.30 & -11.87 &    &   &    &\cr
190 & & dE   &  2 &      & 3 37 29.8 & -18 32 23 & 21.42 &  20.32 & -11.86 &    &   &    &\cr
191 & & dE,N &  3 &      & 3 41 03.3 & -18 04 18 & 20.97 &  20.37 & -11.83 &    &   &    &\cr
192 & & dE/I &  1 &      & 3 40 30.5 & -17 34 08 & 22.16 &  20.49 & -11.80 &    &   &    &\cr
193 & & dE   &  3 &      & 3 38 46.2 & -19 53 15 & 21.72 &  20.34 & -11.79 &    &   &    &\cr
194 & & dE,N &  3 &      & 3 40 37.5 & -18 56 50 & 21.19 &  20.43 & -11.76 &    &   &    &\cr
195 & & dE   &  3 &      & 3 38 06.3 & -19 28 58 & 21.46 &  20.39 & -11.76 &    &   &    &\cr
196 & & dE/I &  2 &      & 3 42 27.2 & -18 46 20 & 21.94 &  20.46 & -11.73 &    &   &    &\cr
197 & & dI   &  3 &      & 3 39 49.3 & -18 20 12 & 21.34 &  20.51 & -11.68 &    &   &    &\cr
198 & & dE/I &  1 &      & 3 40 41.8 & -18 26 11 & 22.71 &  20.55 & -11.62 & 42 & 1 &    20.33 &\cr
199 & & dI/E &  3 &      & 3 43 26.6 & -19 46 26 & 21.77 &  20.56 & -11.61 &    &   &    &\cr
200 & & dE   &  3 &      & 3 43 30.6 & -18 31 39 & 22.05 &  20.60 & -11.59 &    &   &    &\cr
201 & & dE,N &  2 &      & 3 39 42.3 & -18 39 20 & 21.32 &  20.58 & -11.58 & 40 & 3 &    20.26 &\cr
202 & & dE,N &  3 &      & 3 41 19.3 & -18 16 58 & 21.47 &  20.62 & -11.57 &    &   &    &\cr
203 & & dE,N &  3 &      & 3 40 45.1 & -18 48 14 & 21.36 &  20.69 & -11.52 &    & 4 &    &\cr
204 & & dE/I &  3 &      & 3 42 46.7 & -17 49 07 & 22.84 &  20.80 & -11.51 &    &   &    &\cr
205 & & dE   &  3 &      & 3 42 04.6 & -17 31 55 & 22.61 &  20.79 & -11.49 &    &   &    &\cr
206 & & dE/I &  3 &      & 3 41 12.5 & -18 46 10 & 22.87 &  20.73 & -11.48 &    &   &    &\cr
207 & & dE/I &  3 &      & 3 40 00.0 & -18 23 46 & 21.44 &  20.75 & -11.44 & 43 & 3 &    20.44 &\cr
208 & & dE   &  3 &      & 3 41 09.9 & -17 45 06 & 21.45 &  20.82 & -11.41 &    &   &    &\cr
209 & & dE   &  1 &      & 3 37 51.1 & -18 21 41 & 22.08 &  20.81 & -11.38 &    &   &    &\cr      
210 & & dE,N &  3 &      & 3 40 45.2 & -18 48 12 & 21.28 &  20.84 & -11.37 &    &   &    &\cr      
211 & & dE   &  1 &      & 3 41 53.6 & -18 26 32 & 23.63 &  20.82 & -11.37 &    &   &    &\cr      
212 & & dE   &  2 &      & 3 39 29.7 & -18 12 52 & 21.66 &  20.84 & -11.36 &    &   &    &\cr      
213 & & dI   &  3 &      & 3 43 12.2 & -19 27 24 & 22.74 &  20.82 & -11.32 &    &   &    &\cr      
214 & & dE,N &  3 &      & 3 41 33.2 & -17 56 32 & 21.89 &  20.89 & -11.31 &    &   &    &\cr      
215 & & dE/I &  2 &      & 3 34 11.8 & -18 14 21 & 22.53 &  20.85 & -11.31 &    &   &    &\cr      
216 & & dE   &  3 &      & 3 41 23.1 & -18 44 21 & 23.45 &  20.91 & -11.30 &    &   &    &\cr      
217 & & dE   &  3 &      & 3 39 20.9 & -18 15 29 & 21.80 &  20.92 & -11.27 &    &   &    &\cr      
218 & & dI   &  3*&      & 3 39 36.3 & -18 34 45 & 21.62 &  20.93 & -11.17 & 50 & 3 &    20.99   &\cr
219 & & dE/I &  2 &      & 3 40 11.1 & -18 19 31 & 21.62 &  21.01 & -11.17 &    &   &    &\cr      
220 & & dE/I &  3 &      & 3 47 50.5 & -19 58 01 & 22.13 &  21.09 & -11.15 &    &   &    &\cr      
\cr
\noalign{\smallskip \hrule}
\noalign{\smallskip}\cr}}$$}
\end{table*}

\begin{table*}
{\vskip 0.55mm} {$$\vbox{ \halign {\hfil #\hfil && \quad \hfil #\hfil \cr
\noalign{\hrule \medskip}
ID  & Name &  Type & Rating & $V_h$ & $\alpha$ (J2000) & $\delta$ (J2000) & $R_{300}$ & $R$ & $M_R$ & ID$_{\rm TT}$ & Rat$_{\rm TT}$ & $R_{\rm TT}$
                      &\cr
    &            &      &    &  km/s    &  &             &        &        &         &    &   &          &\cr
\noalign{\smallskip \hrule \smallskip}
\cr
221 & & dE,N &  1 &      & 3 38 59.5 & -18 27 27 & 22.76 &  20.95 & -11.14 & 45 & 1 &    20.54 &\cr
222 & & dE   &  2 &      & 3 37 08.4 & -18 10 10 & 22.18 &  21.04 & -11.13 &    &   &    &\cr      
223 & & dE   &  3 &      & 3 38 05.5 & -18 50 25 & 22.48 &  21.07 & -11.13 &    &   &    &\cr      
224 & & dE   &  2 &      & 3 40 25.5 & -18 37 57 & 22.56 &  21.06 & -11.12 & 46 & 2 &    20.75 &\cr
225 & & dE   &  2 &      & 3 39 06.1 & -18 19 45 & 21.98 &  21.08 & -11.10 &    &   &    &\cr
226 & & dE,N &  2 &      & 3 37 24.9 & -18 41 37 & 21.93 &  21.09 & -11.09 &    &   &    &\cr
227 & & dE,N &  2 &      & 3 39 53.1 & -17 57 32 & 22.45 &  21.12 & -11.07 &    &   &    &\cr
228 & & dE/I &  2 &      & 3 37 51.3 & -18 21 39 & 22.07 &  21.12 & -11.07 &    &   &    &\cr
229 & & dE/I &  3 &      & 3 45 41.2 & -18 29 09 & 22.27 &  21.22 & -11.06 &    &   &    &\cr
230 & & dE   &  2 &      & 3 44 26.2 & -17 54 38 & 21.75 &  21.25 & -11.04 &    &   &    &\cr
231 & & dE/I &  3 &      & 3 43 39.1 & -18 57 05 & 22.40 &  21.22 & -11.01 &    &   &    &\cr
232 & & dE/I &  3 &      & 3 47 00.5 & -17 47 32 & 22.47 &  21.21 & -11.01 &    &   &    &\cr
233 & & dE/I &  3 &      & 3 35 08.3 & -18 00 18 & 22.00 &  21.23 & -10.97 &    &   &    &\cr
234 & & dE,N &  3 &      & 3 41 04.4 & -17 52 32 & 22.86 &  21.25 & -10.95 &    &   &    &\cr
235 & & dE   &  3 &      & 3 35 18.5 & -17 48 36 & 21.51 &  21.32 & -10.89 &    &   &    &\cr
236 & & dE,N &  2 &      & 3 38 04.0 & -18 53 43 & 22.28 &  21.33 & -10.85 &    &   &    &\cr
237 & & dE/I &  3 &      & 3 40 52.3 & -19 14 14 & 22.18 &  21.35 & -10.85 &    &   &    &\cr
238 & & dE/I &  2 &      & 3 40 37.9 & -18 16 13 & 22.04 &  21.34 & -10.83 &    &   &    &\cr
239 & & dE/I &  3 &      & 3 40 22.7 & -18 37 02 & 22.51 &  21.38 & -10.80 &    &   &    &\cr
240 & & dE,N &  3 &      & 3 34 16.7 & -18 32 32 & 21.38 &  21.35 & -10.78 &    &   &    &\cr
241 & & dI   &  3*&      & 3 40 16.1 & -18 41 35 & 23.09 &  21.42 & -10.76 & 48 & 2 &    21.20 &\cr
242 & & dE/I &  3 &      & 3 40 06.6 & -18 24 42 & 21.85 &  21.44 & -10.74 & 47 & 3 &    21.04 &\cr
243 & & dI   &  2 &      & 3 41 53.0 & -18 14 49 & 22.74 &  21.45 & -10.73 &    &   &    &\cr
244 & & dE/I &  3 &      & 3 35 23.3 & -19 46 44 & 21.96 &  21.40 & -10.69 &    &   &    &\cr
245 & & dI   &  3*&      & 3 40 44.0 & -18 44 40 & 22.25 &  21.51 & -10.69 & 53 & 2 &    21.51 &\cr
246 & & dI   &  2 &      & 3 40 56.6 & -18 30 40 & 22.37 &  21.48 & -10.68 & 49 & 2 &    21.26 &\cr
247 & & dI   &  3 &      & 3 42 30.5 & -19 52 18 & 22.83 &  21.53 & -10.64 &    &   &    &\cr
248 & & dE   &  3 &      & 3 40 50.7 & -17 45 42 & 21.70 &  21.60 & -10.63 &    &   &    &\cr
249 & & dE/I &  3 &      & 3 40 27.3 & -18 42 30 & 21.86 &  21.57 & -10.62 & 56 & 3 &    21.68 &\cr
250 & & dE/I &  3 &      & 3 41 03.3 & -17 50 33 & 22.47 &  21.64 & -10.57 &    &   &    &\cr
251 & & dE   &  3 &      & 3 42 02.2 & -18 18 20 & 22.27 &  21.61 & -10.56 &    &   &    &\cr
252 & & dE/I &  2 &      & 3 40 48.8 & -18 30 27 & 22.79 &  21.65 & -10.51 & 52 & 2 &    21.46 &\cr
253 & & dE   &  2 &      & 3 38 12.7 & -18 13 43 & 22.35 &  21.67 & -10.50 &    &   &    &\cr
254 & & dE   &  2 &      & 3 39 23.3 & -18 33 16 & 22.41 &  21.70 & -10.46 &    & 4 &    &\cr
255 & & dI   &  3 &      & 3 39 57.3 & -18 38 43 & 21.85 &  21.71 & -10.46 & 51 & 3 &    21.33 &\cr
256 & & dE/I &  3 &      & 3 33 35.9 & -18 27 34 & 22.62 &  21.69 & -10.42 &    &   &    &\cr
257 & & dE/I &  2 &      & 3 41 09.1 & -19 05 50 & 22.67 &  21.82 & -10.38 &    &   &    &\cr
258 & & dE/I &  3 &      & 3 36 24.3 & -18 44 41 & 22.86 &  21.76 & -10.37 &    &   &    &\cr
259 & & dE/I &  3 &      & 3 42 24.0 & -18 54 19 & 22.46 &  21.90 & -10.30 &    &   &    &\cr
260 & & dE,N &  3 &      & 3 40 49.0 & -18 48 35 & 22.27 &  21.93 & -10.28 &    &   &    &\cr
261 & & dI   &  3*&      & 3 39 11.3 & -18 34 11 & 22.69 &  21.96 & -10.20 & 57 & 3 &    21.94 &\cr
262 & & dE   &  3 &      & 3 41 27.1 & -18 46 09 & 22.47 &  22.06 & -10.15 & 54 & 3 &    21.53 &\cr
263 & & dE/I &  3 &      & 3 39 50.4 & -18 35 59 & 22.39 &  22.06 & -10.11 &    & 4 &    &\cr
264 & & dE   &  3 &      & 3 41 08.7 & -18 46 22 & 23.39 &  22.25 &  -9.96 &    &   &    &\cr
265 & & dE/I &  3 &      & 3 40 32.1 & -18 44 43 & 22.45 &  22.25 &  -9.95 &    & 4 &    &\cr
266 & & dE/I &  2 &      & 3 41 06.0 & -19 05 37 & 22.71 &  22.33 &  -9.89 &    &   &    &\cr
267 & & dI   &    &      & 3 39 53.5 & -18 27 54 &       &        &        & 55 & 2 &    21.58 &\cr
268 & & dI   &    &      & 3 38 25.2 & -18 33 51 &       &        &        & 58 & 2 &    22.00 &\cr
\cr
\noalign{\smallskip \hrule}
\noalign{\smallskip}\cr}}$$}
\end{table*}

\section{Results}

\subsection{Spatial correlations}

The spatial distribution of candidate members of the
NGC~1407 Group are shown in Figure 5.  
It is apparent that the
composition is overwhelmingly early-type: 228 have type Sa or earlier 
($T \leq 1$)
and 32 are later types ($T \geq 2$).

\begin{figure*}
\begin{center}
\resizebox{7in}{!}{\includegraphics{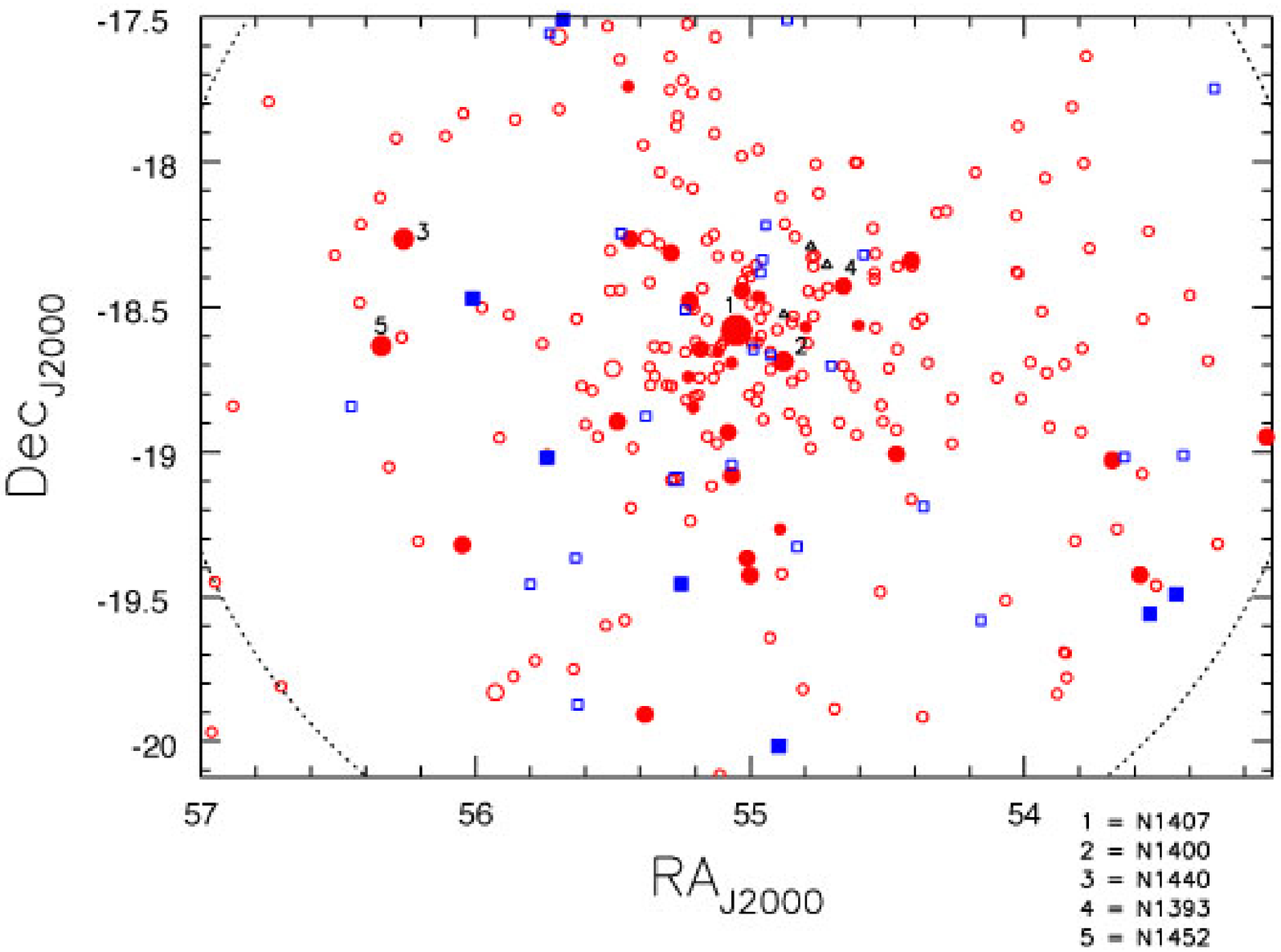}}
\caption{
Positions of candidate NGC~1407 Group members.
Circles represent early-type galaxies and boxes represent late-type galaxies.
Galaxies with filled symbols have their membership confirmed with a velocity.
Large symbols represent galaxies with $M_R < -17$.
The five most luminous galaxies are identified.
Three galaxies in the background within the Southern Wall are
indicated by small triangles.  The outer dotted circle has a radius of 
2~degrees.}
\end{center}
\end{figure*}

There is a strong concentration of early-type
galaxies
toward the center of the group.  The combination of the dominance of
early types and the central concentration is evidence
that the vast majority of the candidates are truly associated with the
group.
There is a caveat: the projected locations of the three S0 galaxies at 
4200~km~s$^{-1}$
indicated by the small open triangles are very
near the centre of the group.  They are known to be background but
their neighbors might enter our sample.
The possibility is difficult to quantify but this structure 
is small
so we expect that the contamination is also small.

We present two-point correlations for various galaxy subsamples in Figure 6.  
Following
MTT05, the correlations are normalized by comparison with Monte Carlo random
catalogs.  We do not have a fair sample of the Universe so we give
attention only to {\it relative} variations in correlations.

Very strong correlations are seen in the upper left panel of Fig.~6.  The points
show the correlation of all types of galaxies that are either known
to be members on the basis of velocities or are rated 1 and 2, which we
suspect are almost certain to be members.  The solid curve gives the correlation
if the sample is restricted to only early-type galaxies ($T \leq 1$).
These early-type galaxies are slightly more strongly correlated.  We do not take
the poorer correlation of late types in this sample as
evidence that
they are not in the group (a half dozen have memberships confirmed by 
velocities); it would not be surprising if late-types typically live
further out from the group center than the early types.

The upper right panel of Fig.~6 shows the correlation of early-type galaxies rated 3 (plausible
members) in comparison with the fiducial curve of early-type galaxies with
known velocities or rated 0--2.  From the weaker correlation one infers that
a fraction of those early-type objects rated 3 are not real members.  
The dotted curve
illustrates the correlation found if the fiducial correlation is contaminated
with a 20\% random population.  We is conclude that $80 \pm 10\%$ of the
early-type galaxies rated 3 are group members.

The correlation of only late-type ($T \geq 2$) 
galaxies seen in the lower left panel of Fig.~6 is much
weaker.  However we do not take this as evidence against group membership
because a majority of this sub-sample have velocities or are
rated 1 or 2. 
We have not shown the correlation for late-type galaxies rated 3 only
because there are just 12 such objects.

The lower right panel of
Fig.~6 shows the correlation among only the 34 brightest galaxies.
Almost all these galaxies are known to be group members on the basis
of velocity measurements.  Yet the 2--point correlation is much weaker than
for the fiducial sample!  The two brightest galaxies, NGC~1407 and
NGC~1400, are very close to the  group center.  On the contrary, the next
brightest galaxies, in the range $0.04 - 1 L^{\ast}$ are relatively
dispersed compared with the fiducial population numerically dominated by
early-type dwarfs.

\begin{figure}
\begin{center}
\vskip-4mm
\psfig{file=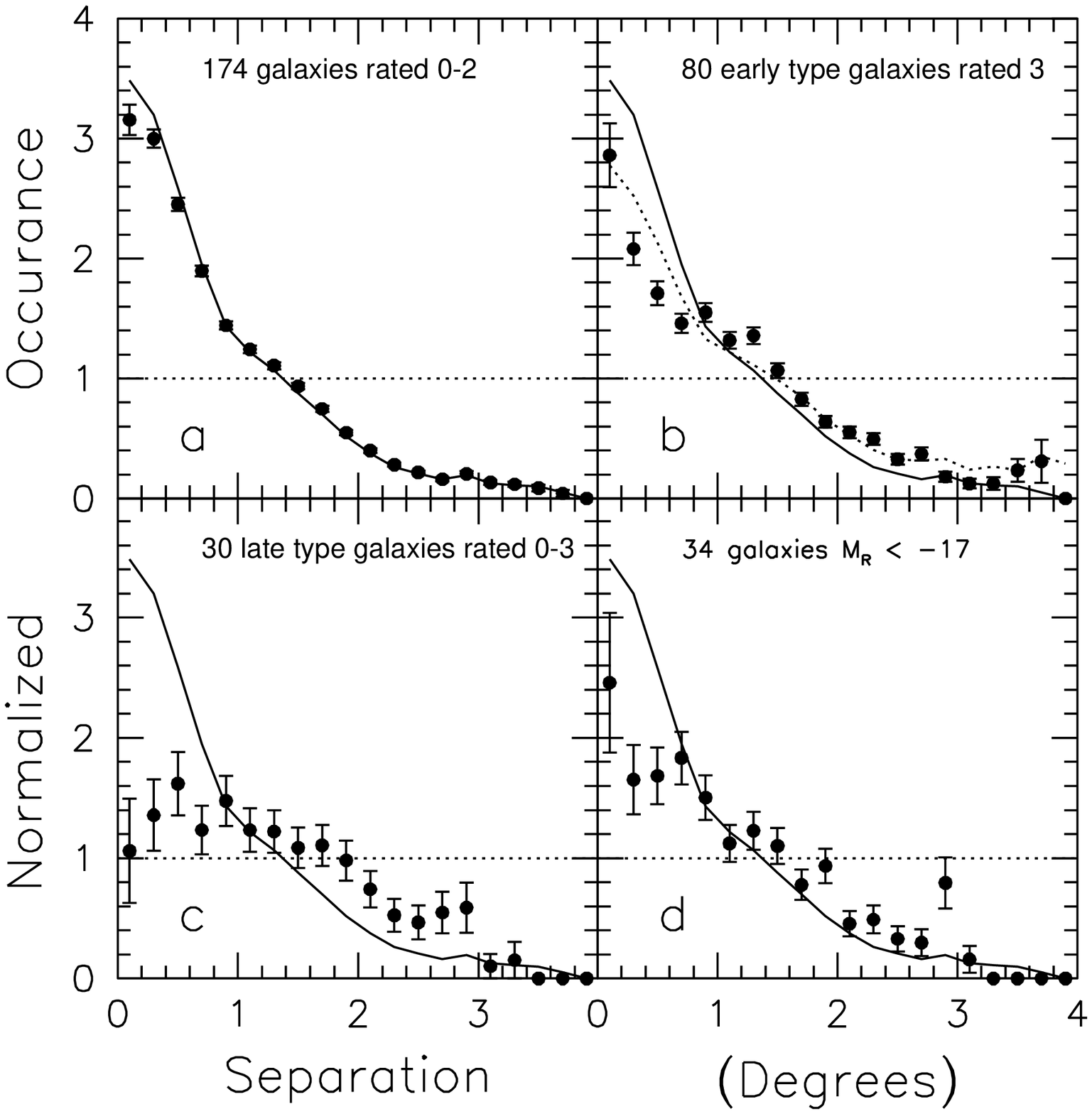, width=8.65cm}
\end{center}
\vskip-3mm
\caption{
Two--point correlation functions.  Pairwise separations are compared against
Monte Carlo random distributions for several subsets of galaxies in the
CFHT survey region around NGC~1407.  In panel $a$, the points with error
bars represents the correlation found among the 172 galaxies either with
velocities that identify them as group members or rated 1 and 2, hence
almost certainly group members.  The solid line demonstrates the correlation
with the 154 galaxies typed Sa or earlier and with a velocity or rated 1 and
2.  This highly correlated sample provides a reference which is carried
over to all the panels.  In panel $b$ one sees the correlation among 80
early types rated 3.  The slightly weaker correlation can be matched by
mixing the reference correlation with a 20\% addition of randomly distributed
objects.  Panels $c$ and $d$ show the weaker correlations that are found
among 30 galaxies typed later than Sa and among 34 galaxies brighter than
$M_R=-17$, respectively.  In the case of the brighter galaxies, most of
the systems have measured velocities that confirm membership so the
weaker correlation indicates this population is more dispersed in the group.
The weaker correlation among late types may at least in part have a similar
origin. 
}
\end{figure}

We draw the following conclusions about our subjective ratings from the spatial correlations.
There is no reliable information that contradicts the conclusions of MTT05 drawn
from the NGC~5846 Group analysis that systems rated 1 or 2 are very likely to be group members
and that systems rated 4 are very likely to be background.
We find that the background
contamination within the sample of early-type objects rated 3 is a bit 
smaller in the NGC~1407
Group ($\sim 20\%$) than in the NGC~5846 Group ($30-50\%$ depending on
luminosity).  
The late-type dwarfs are more weakly correlated but we cannot decide
whether that is due to contamination or a consequence of an intrinsically 
weaker correlation
(as could arise if late types arrived more recently in the group).

Our sample from the CFHT survey amounts to 260 galaxies 
(9 additional galaxies listed in
Table 2 are from the deeper but limited Subaru survey and are not considered
part of the statistical sample).  
About 34 additional candidates would be expected
to be found if the CFHT survey were extended to fill the $2^{\circ}$ circle
corresponding to the second turnaround radius described later in this section.
Roughly 47 of the ensemble of
candidates are expected on statistical grounds to be
background contaminants.  
In summary we expect the NGC~1407 to contain $\sim 250 \pm 20$ galaxies 
to the limit of our survey.

\subsection{The distinct radial distributions of dwarfs and giants}

The spatial correlation differences brought to light in Fig.~6 are shown
in another way in Fig.~7.  Here we show the cumulative radial distributions
of dwarf ellipticals ($M_R>-17$) and of all bright group members 
($M_R \le -17$).  The luminous galaxies are less concentrated toward
the center of the group than the dwarf galaxies; these two populations have
50 percentile radii of $0.79^{\circ} = 350$~kpc and
$0.55^{\circ} = 240$~kpc respectively.
Miles et al.~(2004) draw attention
to an apparent dip in the luminosity function in X-ray dim groups.
They ascribe this feature to dynamical evolution, with
intermediate-mass systems merging with the central massive system on 
shorter timescales than lower-mass systems.
A similar process may be operating here 
and may generate the observed phenomenon.
The radial scale of the effect manifested
in Fig.~7 is $\sim 1^{\circ}$, roughly half the virial radius.  Jones
et al.~(2003) point out that orbital decay by dynamical friction will
cause an $L^{\star}$ galaxy to fall from this radius to the center of 
an NGC~1407-like group in a Hubble time.

\begin{figure}
\begin{center}
\vskip-4mm
\psfig{file=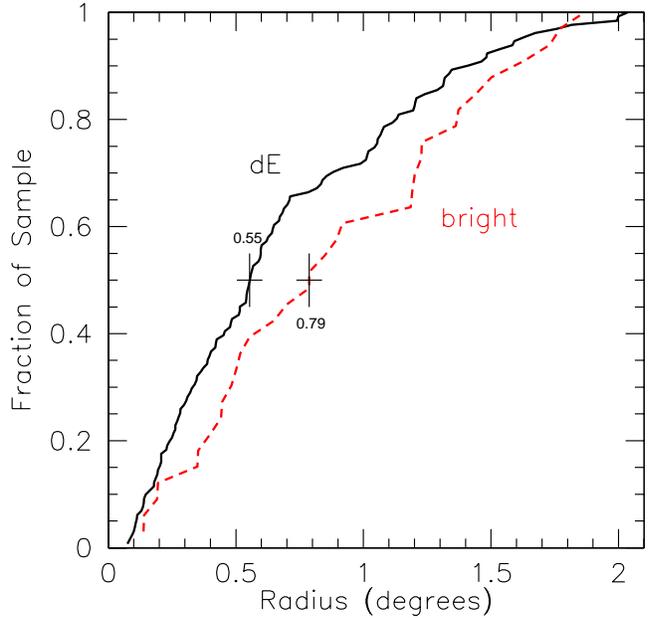, width=8.65cm}
\end{center}
\vskip-3mm
\caption{
Cumulative radial distributions for the NGC~1407 Group.  The solid curve at left 
shows the
cumulative radial distribution of all (131) early types ($T \le 1$) rated
1 or 2 fainter than $M_R=-17$.  The dashed curve at the right shows the
cumulative radial distribution of all (33) galaxies brighter than $M_R=-17$.
}
\end{figure}

\subsection{The second turnaround radius}

In the case of the NGC~5846 Group, MTT05 claimed to identify a drop in the
radial density distribution and the radial velocity distribution at a
radius of 840~kpc which was identified as the caustic of second turnaround
(Bertschinger 1985).  
The NGC~5846 analysis benefited from the availability
of a complete sample of velocities down to $M_R \sim -15$ across an extent
much larger than the group from the Sloan Digital Sky Survey.  There were
three times more velocities available within that group and the domain beyond
the group could be characterized.

With collisionless spherical collapse, the surface of second turnaround today
corresponds to the same density level for all collapsed objects.  Comparing
the second turnaround radii $r_{2t,i}$ of two groups or clusters of mass
$M_i$, $i=1,2$:
\begin{equation}
M_1/M_2 = (r_{2t,1}/r_{2t,2})^3.
\end{equation}
Using the NGC~5846 calibration, then $r_{2t}$ for a group of mass $M_{12}$ in 
mass units of $10^{12} M_{\odot}$ is
\begin{equation}
r_{2t} = 0.193 (M_{12})^{1/3}~{\rm Mpc}.
\end{equation}
Using the virial approximation, with $M \propto \sigma_v^2 r_{2t}$,
\begin{equation}
r_{2t} = \sigma_v / 390 ~{\rm Mpc}.
\end{equation}
Equations 4 and 5 
give alternate values for $r_{2t}^{N1407}$
of 810~kpc ($1.84^{\circ}$) and 990~kpc ($2.27^{\circ}$).  
An approximate value for the second turnaround radius is therefore $2.0^{\circ}$.
On Figure 8, we show where this radius falls on a plot of the surface density of galaxies.
This $r_{2t}$ estimate
reasonably approximates the
boundary of the group. It extends modestly outside the
survey region.

\begin{figure}
\begin{center}
\vskip-4mm
\psfig{file=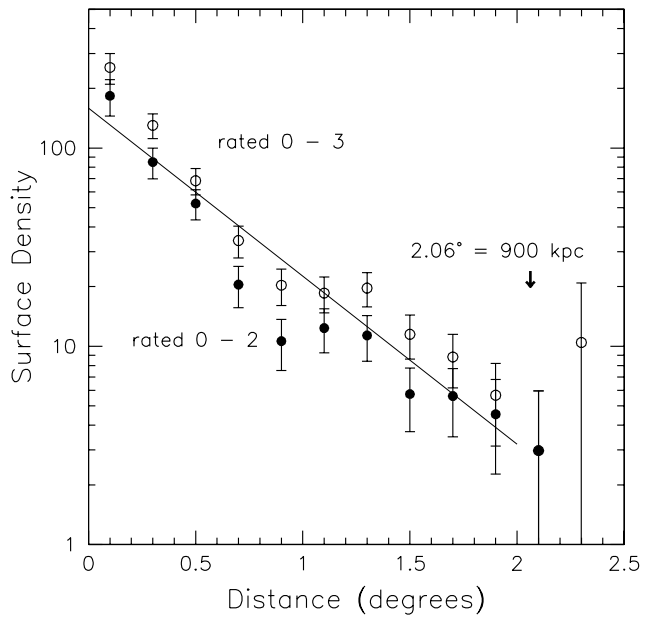, width=8.65cm}
\end{center}
\vskip-3mm
\caption{
The variation of surface density of galaxies with radius.  Filled symbols 
refer to systems with
velocities or rated 1 and 2.  Open symbols include the contribution from systems rated 3.
Corrections for the restricted boundary of the survey are included.
These are significant as radii greater than
1 degree and result in extremely large uncertainties for the 3 points at largest radii.
The straight line represents a fit to the average of the pairs of points
inside 2 degrees and has a slope of 0.85, implying a fall-off in the
volume density with a slope of 1.85.  The inferred radius of second turnaround
is indicated at $2.06^{\circ} = 900$~kpc.}
\end{figure}

\subsection{Galaxy scaling relations}

The correlation between central surface brightness and absolute
luminosity is shown in Figure~9.  The measure of central surface brightness
is $R_{300}$, the luminosity in a $300~{\rm pc} = 2.5^{\prime\prime}$
radius aperture.
The equivalent plot in MTT05 looks similar. 
For the galaxies in the sample, lower-luminosity galaxies have a 
systematically lower central surface brightness.  Equivalently, as shown by 
MTT05, lower luminosity galaxies have lower surface brightnesses within
the `effective' radius containing half the light.
Early type galaxies tend to lie above late type galaxies
of the same magnitude.
There is a hint of the
gap at $R_{300} \sim 16$ seen in MTT05; galaxies with absolute magnitudes
$M_R \sim -19$ have a tendency to 
avoid intermediate central surface brightnesses characterized by 
$R_{300} \sim 16$.

\begin{figure}
\begin{center}
\vskip-4mm
\psfig{file=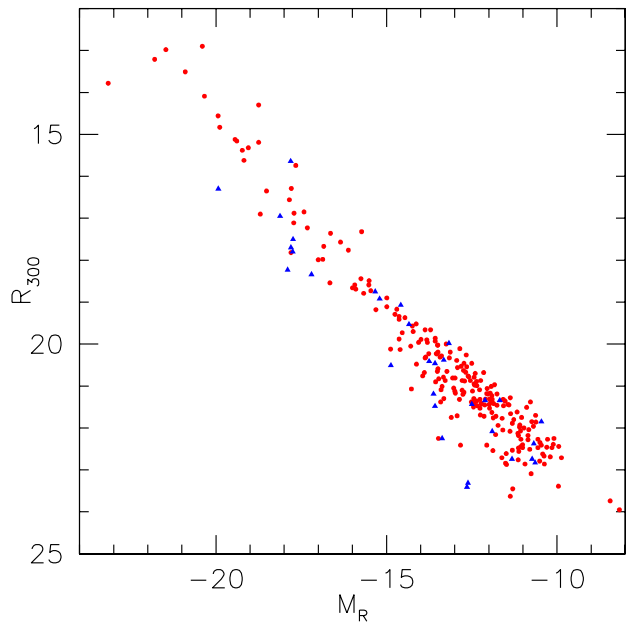, width=8.65cm}
\end{center}
\vskip-3mm
\caption{
Central luminosity vs total luminosity.  The central luminosity is the
$R$-band flux measured within a 300 pc = 2.5 arcsec radius aperture.
Types $T \le 1$ and $T > 1$ are distinguished by circles and triangles
respectively.
}
\end{figure}

\subsection{Normalization of the luminosity function}

In TT02 and MTT05 we introduced a normalization of
the luminosity function which enables us to compare results from different
environments.
The normalization is based on the number
density of galaxies brighter than $M_R = -17$ at a fiducial radius of
200~kpc.  For the NGC~1407 group the normalization is set at 34
galaxies/Mpc$^2$ (see Figure 10).
The two sets of points plotted in Fig.~10 represent the radial distributions of bright
galaxies and the equivalent distribution of the entire sample.  The former
is wanted to establish the luminosity function normalization just discussed.
The latter is dominated by the dwarf population.

\begin{figure}
\begin{center}
\vskip-4mm
\psfig{file=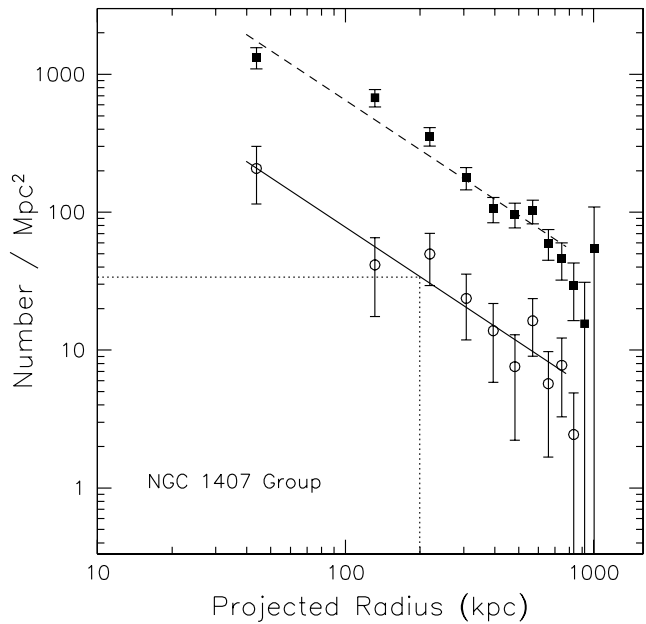, width=8.65cm}
\end{center}
\vskip-3mm
\caption{
The radial distribution of bright galaxies and of all galaxies.  The sample
of galaxies with $M_R < -17$ is indicated by the open symbols and provide
the surface density calibration at 200~kpc used to normalize the luminosity
function.  The filled symbols indicate the distribution of the entire
sample.  The solid line represents a fit to the bright sample and the
dashed line is this same line shifted vertically so as to be appropriate to the entire sample.
}
\end{figure}

\subsection{Luminosity function}

The group luminosity function
is shown in Figure~11.  The small filled and open circles show the range in
possibilities if {\it no} objects rated 3 are members or {\it all}
objects rated 3 are members, respectively.  The squares with
error bars are determined on the assumption that, brighter than $-13.4$, 
70\% of the objects rated 3 are members, while fainter than $-13.4$ that 
50\% of early types are members but that none of later types are members.
(the same rules determined for the NGC~5846 Group sample by MTT05).
The turndown in the final two points is presumably due to incompleteness.  

There is a dearth of luminous systems in the group after the brightest
system, NGC~1407, at $4 L^{\star}$.  The second
brightest galaxy, NGC~1400, has a luminosity that is $\sim L^{\star}$.
The dwarf-to-giant ratio introduced by TT02 is defined as the number of
galaxies with $-17 < M_R < -11$ over the number of galaxies with 
$M_R < -17$.  For the NGC~1407 Group, that ratio is 
$200/31 = 6.5 \pm 1.3$ where the uncertainty includes estimates of the
incompleteness near the $M_R = -11$ limit and membership uncertainties.
This ratio is comparable with the value of $7.3 \pm 0.7$ found by MTT05
for the NGC~5846 Group but significantly larger than the ratio 
$2.6 \pm 0.9$ found for 4 other groups in the Local Supecluster by TT02
(Coma~I, Leo, NGC~1023, Ursa Major).

\begin{figure}
\begin{center}
\vskip-4mm
\psfig{file=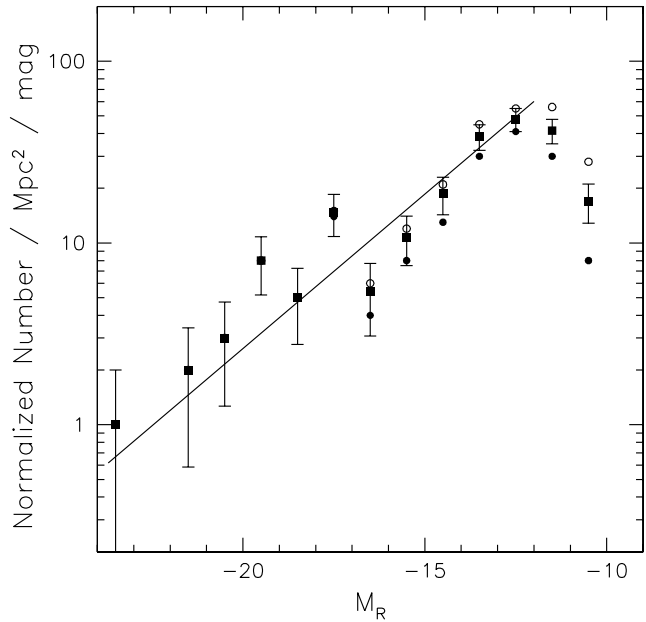, width=8.65cm}
\end{center}
\vskip-3mm
\caption{
NGC~1407 Group luminosity function.  
Filled squares give normalized numbers per magnitude of galaxies confirmed
as group members with velocities plus those with membership ratings 1 and 2,
plus fractions of those with membership rating 3 as discussed in the text.
Small filled circles: number with
measured velocities or with membership ratings 1 and 2.  Small open circles:
also include galaxies with membership rating 3.  The true luminosity function
is expected to be bracketed by these extremes.  
Error bars reflect $\sqrt N$ statistical variances.
The straight line is a linear fit to the squares with error bars for
$M_R < -12$.
The slope is equivalent to a faint end Schechter function slope of
$\alpha = -1.43 \pm 0.05$.
}
\end{figure}

\begin{figure}
\begin{center}
\vskip-4mm
\psfig{file=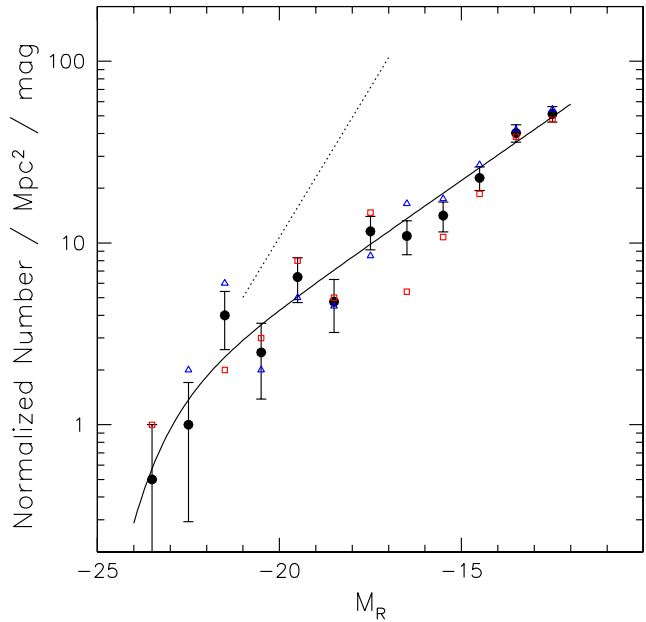, width=8.65cm}
\end{center}
\vskip-3mm
\caption{
Luminosity function for the combined NGC~1407 and NGC~5846 groups.
The large filled symbols with error bars show the combined data.
The small squares and triangles give the data for NGC~1407 and NGC~5846
respectively.  The solid curve illustrates a Schechter function fit to the
combined data.  
The dotted line shows the slope anticipated for the 
mass function in a $\Lambda CDM$ cosmology with arbitrary normalization.
}
\end{figure}

\begin{figure}
\begin{center}
\vskip-4mm
\psfig{file=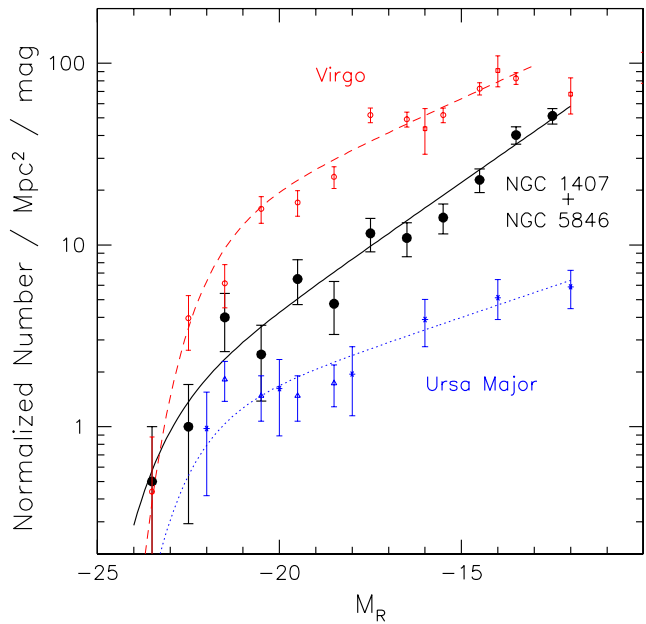, width=8.65cm}
\end{center}
\vskip-3mm
\caption{
Luminosity function comparisons with the Virgo and Ursa Major clusters.  
The luminosity 
function found from the combined NGC~1407 and NGC~5846 data sets is 
compared with the luminosity function for the Virgo Cluster reported in
TT02 and for the U~Ma Cluster reported by Trentham et al (2001).  
Normalizations follow the same procedures.
The solid curve is the Schechter function fit to the combined
NGC~1407 and NGC~5846 material with Schechter function parameters 
$M^{\star}=-23.6$ and $\alpha=-1.35$.  The dashed curve shows the
Schechter function fit to the Virgo data with $M^{\star}=-22.2$ and
$\alpha=-1.23$.   The dotted curve show such a fit to the U~Ma data with
$M^{\star}=-22.7$ and $\alpha=-1.17$. 
}
\end{figure}

\begin{figure}
\begin{center}
\vskip-4mm
\psfig{file=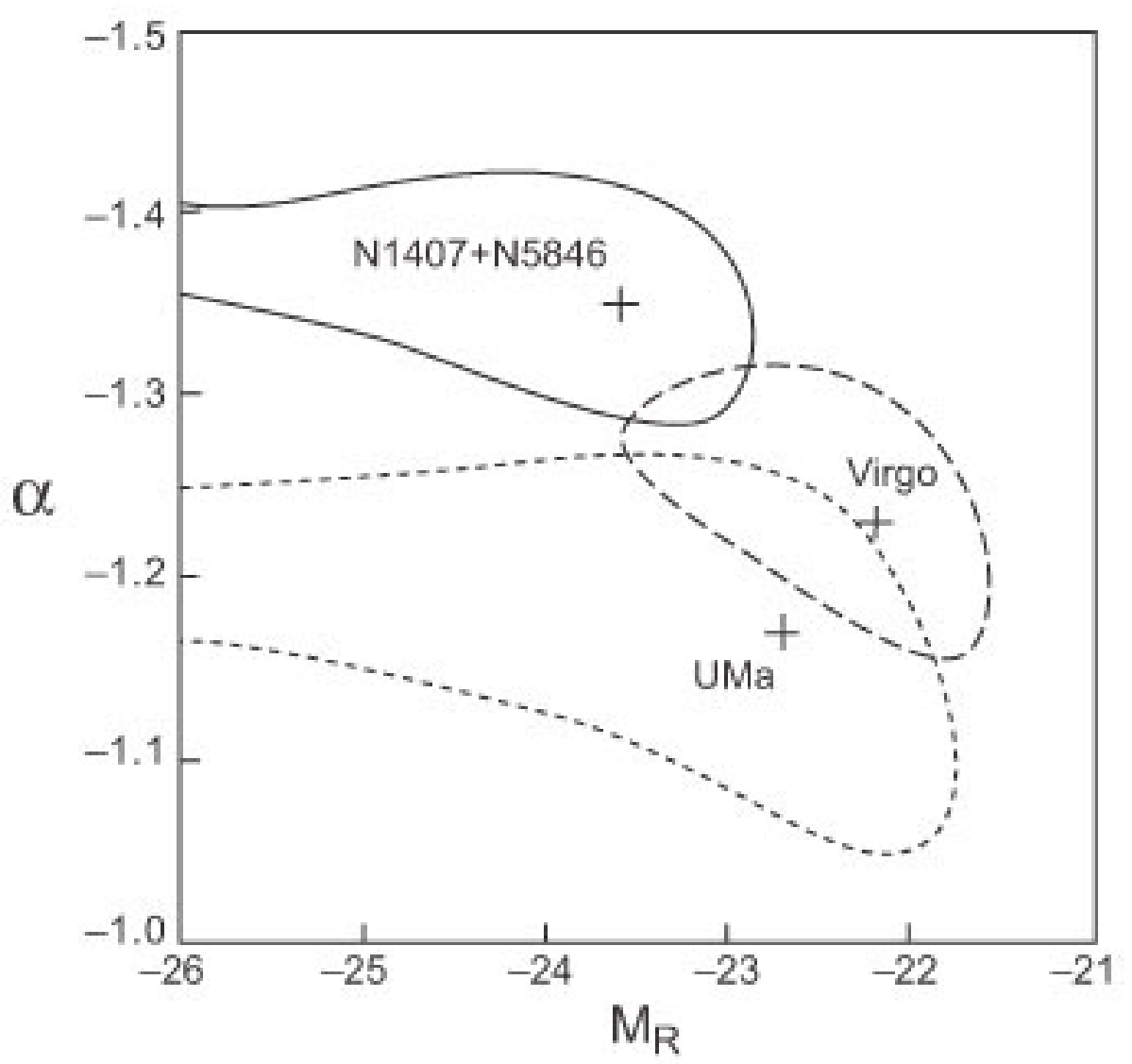, width=8.65cm}
\end{center}
\vskip-3mm
\caption{
$\chi^2$ constraints on parameters for Schechter function fits.  
Best $\chi^2$ values lie at the
crosses.  Contours are 95\% probability.  Solid contour: combined 
NGC~1407/5846 groups.  Dashed contour: Virgo Cluster.  Dotted contour: 
Ursa Major Cluster.  The contours open at high luminosity for 
NGC~1407/5846 and U Ma reveal that linear
fits in these cases have $\chi^2$ significance within $2 \sigma$ of the
best Schechter function fits. 
}
\end{figure}

In MTT05 and TT02 we attempted to fit separate luminosity functions to the 
bright and faint ends but
such an exercise is not warranted here, because of small number statistics
at the bright end.
The straight line seen in
Fig.~11 is a fit to the data with error bars.
The fit conforms to the
constraint of zero objects in the bin at $-23 < M_R < -22$ and then one
(NGC~1407) in the next brighter bin.  The slope of this line corresponds to the faint-end slope in the
Schechter (1976) formulation of $\alpha = -1.43$.  The
usual exponential cutoff at the high luminosity end cannot be defined 
because there are
so few objects above $M_R = -20$ and one very luminous system.

The discussion of the luminosity function is most meaningfully extended
by amalgamating the NGC~1407 and (MTT05) NGC~5846 results.  This is a
particularly useful exercise because (a) the two groups are at 
similar distance, (b) observations of both were with the same CFHT
camera for the same exposure times and reductions for both were analyzed with the
same algorithms, and (c) the groups are very similar in dimensions,
velocity dispersions (hence masses) and morphological compositions.
The luminosity functions of the two groups are shown in Figure~12,
separately through the small symbols and jointly through the large symbols.
The two groups are seen to follow the same luminosity distribution
without a significant offset in number surface density (after the
normalization computed in the case of NGC~1407 using Fig.~10 and in the case of NGC~5846
using Fig.~12 in MTT05).  The average luminosity function of the two groups is
well-fit with a Schechter function
(the solid line in Fig.~12.) with a faint end slope 
$\alpha = -1.35$.  The bright-end exponential
cutoff is characterized by $M_R^{\star} = -23.6$.  This unusually
bright cutoff can be attributed to the evolved states of the two groups.
The dominant galaxies are unusually bright.

The dotted line in Figure~12 demonstrates the slope of the 
mass function
given by hierarchical clustering in the standard $\Lambda CDM$ model.
Clearly, the luminosity function is shallower.

Other comparisons with the combined NGC~1407/5846 luminosity function 
are made in Figure~13.  The data defining luminosity functions found in the
Virgo Cluster by TT02 and in the Ursa Major Cluster by Trentham et al.
(2001) are superposed, scaled by the same normalization procedure.
The curves are Schechter function fits characterized
by a faint end slope parameter $\alpha$ and a bright end exponential
break at $M^{\star}$.  The coupling between these parameters is seen in 
Figure~14.  It is seen that the bright end break does not quite have
$2 \sigma$ significance for either the combined NGC~1407/5846 or U~Ma
samples.  It may seem that the Virgo and NGC~1407/5846 samples have different
slopes but the apparent difference is largely a consequence of the coupling
between the parameters $M^{\star}$ and $\alpha$ and the clear requirement
of a lower $M^{\star}$ with the Virgo data.

\section{Discussion and Conclusion}

The NGC~1407 Group is overwhelmingly an environment of galaxies deficient
in cold gas and young stars.  This point is emphasized by Figure~15,
a plot updated from TT02 and MTT05.  The environment is even more 
extreme in this respect than in the NGC~5846 Group or the core of the Virgo
Cluster.

\begin{figure}
\begin{center}
\vskip-4mm
\psfig{file=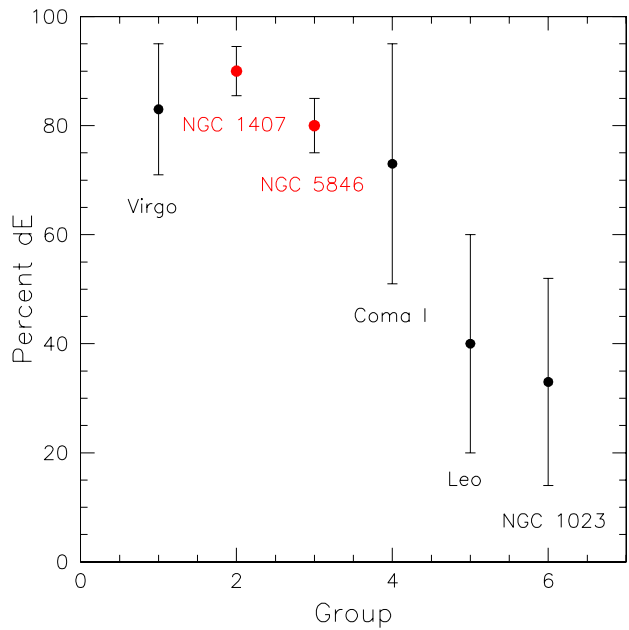, width=8.65cm}
\end{center}
\vskip-3mm
\caption{
Percent of dwarfs that are early type.  Galaxies with luminosities
in the range $-17 < M_R < -11$ are split between early (dE, dE/I)
and late types (VLSB, dI, dI/E).  In the case of the NGC~1407 Group
$90 \pm 5\%$ are early type.  This fraction is compared with the values
found by MTT05 and TT02 for other groups.}
\end{figure}

Here is a review of properties of the NGC~1407 Group that collectively
suggest that the group is at an advanced state of dynamical evolution.

\noindent$\bullet$ 
The galaxy population is almost entirely Population II ellipsoids.

\noindent$\bullet$
The group is dominated by a single luminous galaxy $1.4^m$ brighter than
the second brightest galaxy.

\noindent$\bullet$ 
That second brightest galaxy, NGC 1400, has an incredible
differential motion of $-1072$~km~s$^{-1}$ with respect to the group
mean.

\noindent$\bullet$ 
The bright galaxies are less concentrated to the center of
the group than the dwarfs.

The NGC~1407 Group shares many of the properties of
fossil groups (Ponman et al. 1994).

The depletion of intermediate luminosity systems in an evolved group
is attributed to merging and subsequent growth of the
central dominant galaxy (Jones et al. 2003).  D'Onghia et al. (2005)
have followed N-body/hydrodynamical models of clusters with similar
masses of $\sim 10^{14} M_{\odot}$ and found a correlation between the
magnitude interval between first and second brightest objects and the
epoch when a group assembled 50\% of its final mass.  According to their 
simulations,
a group with the properties found here would have assembled at $z \sim
0.8$.

We now show that this group possesses many dwarf galaxies,
another similarity with fossil groups
(Ponman et al. 1994; Mulchaey and
Zabludoff 1999).
This tells us that the physical process responsible for depleting the 
cluster of intermediate luminosity galaxies does not remove the entire dwarf galaxy population.  This is the primary result of this paper.
In the context of our wider study, the presence of a substantial dwarf elliptical population could well be a ubiquitous feature of dynamically evolved 
environments.

By comparison, the NGC~5846 Group studied by MTT05 appears to be younger.  
The two groups
have comparable dimensions and velocity dispersions, hence total masses.
However, the NGC~5846 group is a dumbbell system with two large ellipticals
of comparable luminosity, each the center of an enhanced dwarf population.
There are more intermediate luminosity galaxies in that group, with 5
$L^{\star}$ galaxies, including two spirals.  The estimated
total number of group members brighter than our survey limits are $\sim 250$ in both
cases.

The large-scale environments of the two groups are somewhat different.  The NGC~5846
group is a knot of early type galaxies in a filament that is otherwise
low density and spiral rich.
The spiral dominated filament probably provides
the source for more recent infall into the NGC~5846 Group.
By contrast, the NGC~1407 Group is
adjacent other knots of early type galaxies.  The larger region
around NGC~1407 appears to be more dynamically evolved than the larger
region around NGC~5846.  In section 2.1 we inferred there is somewhat
more than $10^{14} M_{\odot}$ in the Eridanus region identified by
Tully (1988) as the 51-xx+4 association.  The zero velocity or first
turnaround surface at radius $r_{1t}$
enclosing the bound region of this association can be estimated by scaling
from the Local Group (Tully 2005):
\begin{equation}
r_{1t}^{Eri} \sim r_{1t}^{LG} (M^{Eri}/M^{LG})^{1/3} \sim 4~{\rm Mpc}
\sim 9^{\circ}.
\end{equation}
A circle of this radius projected onto the Eridanus region encloses most
of the galaxies plotted in the left panel of Fig.~2.  This region should
collapse in on itself over the next Hubble time forming an entity
comparable to the Fornax Cluster as seen today.  The NGC~1407 Group
(51-8 in the NBG catalog) appears at the moment to be the dominant mass
component but the group around NGC~1395 and NGC~1398 is slightly more
luminous and contains many more intermediate luminosity galaxies.
The NGC~1407 group may be close to a fossil group today but it will be 
reinvigorated
with new group members in the future.

The very high relative motion of NGC~1400 is a special feature of the
NGC~1407 Group.  The existence of a second highly blueshifted
object, albeit a much less luminous system, shows that the motion of
NGC~1400 is not an unique event.  Both high
velocity systems are very near to NGC~1407 and the center of the group.
This lends additional support for the scenario advocated by Gould (1993):
that there exists a 
very massive
gravitational well around NGC~1407.  
In the context of the alternative model described earlier, 
the second blueshifted galaxy is a companion that is bound to NGC~1400
as both move past the group.

While there are certainly more dwarf galaxies in dense environments than 
in diffuse environments, there still appear to be fewer dwarf galaxies than
low-mass dark matter subhalos (De Lucia et al.~2004).  This tells us that whatever physical 
processes are responsible for the suppression of star formation in low-mass 
galaxies must operate less efficiently in clusters than in the
field or that there are additional processes that operate in clusters 
or regions that evolve into clusters that  
cause the formation of dwarf galaxies. 
More definitive statements can be made when measurements for a large 
sample of groups exhibiting a range of properties are available.

\section*{Acknowledgements}

Support for RBT was provided by the US National Science Foundation
award AST 03-07706.  This research has made use of the NASA/IPAC
Extragalactic Database (NED) which is operated by the Jet Propulsion
Laboratory, Caltech, under agreement with the National Aeronautics and
Space Association.

Imaging observations were with the the Canada-France-Hawaii Telescope
(CFHT) facility MegaPrime/MegaCam, a joint project of
CFHT and CEA/DAPNIA.  CFHT is operated by the National Research Council
of Canada, the Institute National des Sciences de l'Univers of the
Centre National de la Recherche Scientifique of France, and the
University of Hawaii.  Spectroscopic observations were made at the
W.M. Keck Observatory.

Images of the galaxies described in this paper can be obtained from the CFHT cutout service:
http://www.cadc.hia.nrc.gc.ca/cadcbin/cfht/cfhtCutout.

\end{document}